\documentclass[preprint]{aastex}
\usepackage[bookmarks=false]{hyperref}
\usepackage{amssymb, amsmath}
\usepackage{arydshln}
\usepackage{xcolor}
\usepackage{rotating}
\usepackage[normalem]{ulem}
\usepackage{multirow}
\newcommand{\cmnt}[1]{}

\newcommand{\suma}{{\Sigma_a}}

\begin{document}

\title{Brighter-fatter effect in near-infrared detectors -- III. Fourier-domain treatment of flat field correlations and application to {\slshape WFIRST}}

\date{March 12, 2020}
\author{Jenna K.C. Freudenburg$^{1,2}$, Jahmour J. Givans$^{2,3}$, Ami Choi$^{2,3}$, Christopher M. Hirata$^{1,2,3}$, Chris Bennett$^4$, Stephanie Cheung$^4$, Analia Cillis$^4$, Dave Cottingham$^4$, Robert J. Hill$^4$, Jon Mah$^4$, and Lane Meier$^4$}
\affil{$^1$Department of Astronomy, The Ohio State University, 140 West 18th Avenue, Columbus, Ohio, 43210, USA}
\affil{$^2$Center for Cosmology and AstroParticle Physics, The Ohio State University, 191 West Woodruff Avenue, Columbus, Ohio 43210, USA}
\affil{$^3$Department of Physics, The Ohio State University, 191 West Woodruff Avenue, Columbus, Ohio, 43210, USA}
\affil{$^4$ NASA Goddard Space Flight Center, Detector Characterization Laboratory, 8800 Greenbelt Rd., Greenbelt, Maryland 20771, USA}

\email{freudenburg.2@osu.edu}

\begin{abstract}
Weak gravitational lensing has emerged as a leading probe of the growth of cosmic structure. However, the shear signal is very small and accurate measurement depends critically on our ability to understand how non-ideal instrumental effects affect astronomical images. The {\slshape Wide-Field Infrared Survey Telescope} ({\slshape WFIRST}) will fly a focal plane containing 18 Teledyne H4RG-10 near infrared detector arrays, which present different instrument calibration challenges from previous weak lensing observations. Previous work [Paper I: Hirata \& Choi, PASP, 132, 014501 (2020); and Paper II: Choi \& Hirata, PASP, 132, 014502 (2020)] has shown that correlation functions of flat field images, including cross-correlations between different time slices that are enabled by the non-destructive read capability of the infrared detectors, are effective tools for disentangling linear and non-linear inter-pixel capacitance (IPC) and the brighter-fatter effect (BFE). Here we present a Fourier-domain treatment of the flat field correlations, which allows us to expand the previous formalism to all orders in IPC, BFE, and classical non-linearity. We show that biases in simulated flat field analyses in Paper I are greatly reduced through the use of this formalism. We then apply this updated formalism to flat field data from three {\slshape WFIRST} flight candidate detectors, and explore the robustness to variations in the analysis. We find that the BFE is present in all three detectors, and that its contribution to the flat field correlations dominates over the non-linear IPC, in accordance with the results from Paper II on a development detector. The magnitude of the BFE is such that the effective area of a pixel is increased by $(3.54\pm0.03)\times 10^{-7}$ for every electron deposited in a neighboring pixel (SCA 20829, statistical error, not IPC-deconvolved). We compare IPC maps from flat field autocorrelation measurements to those obtained from the single pixel reset method and find a median difference of 0.113\% for SCA 20829. After further diagnosis of this difference, we ascribe it largely to an additional source of cross-talk, the vertical trailing pixel effect, and recommend further work to develop a model for this effect. These results represent a significant step toward calibration of the non-ideal effects in {\slshape WFIRST} detectors.
\end{abstract}

\keywords{instrumentation: detectors}

\section{Introduction}
\label{sec:intro}

One of the most important results in modern cosmology was the discovery using Type Ia supernovae that the expansion of the Universe is accelerating \citep{1998AJ....116.1009R, 1999ApJ...517..565P}. This has motivated a wide range of observational probes using different techniques to measure both the expansion history of the Universe and the growth of cosmic structure, and test whether these are consistent with the simplest model of a cosmological constant and unmodified general relativity \citep[e.g.][]{2006astro.ph..9591A, 2013PhR...530...87W}. Weak gravitational lensing has emerged as one of the key tools in this effort. With recent surveys including the Kilo Degree Survey (KiDS; \citealt{2020A&A...633A..69H}), the Dark Energy Survey (DES; \citealt{2018PhRvD..98d3526A}), and the Hyper Suprime-Cam (HSC; \citealt{2019PASJ...71...43H}) the most powerful tool for growth of structure at low redshift. Even more ambitious surveys are planned for the future, including the Legacy Survey of Space and Time (LSST\footnote{\url{http://www.lsst.org}}) at the ground-based Vera Rubin Observatory, and the {\slshape Euclid}\,\footnote{\url{http://sci.esa.int/euclid}} \citep{2011arXiv1110.3193L} and {\slshape Wide Field Infrared Survey Telescope} ({\slshape WFIRST}\,\footnote{\url{http://wfirst.gsfc.nasa.gov}}; \citealt{2015arXiv150303757S}) satellites. With the larger data sets from these surveys, tight control of observational and astrophysical systematics will be of central importance.

While the actual observing program for {\slshape WFIRST} has not yet been decided, the {\slshape WFIRST} reference program includes a 2000 deg$^2$ survey in four near infrared (NIR) bands covering 0.9--2.0 $\mu$m, with the image quality (point spread function $\sim 0.13$ arcsec half light radius), NIR depth (26.2--26.95 mag AB at 5$\sigma$, depending on the band), and stability available from space \citep{2018arXiv180403628D, 2019BAAS...51c.341D, 2019arXiv190205569A}. In order to make precise measurements of the already small weak lensing signal, shear measurement systematics must be kept under control at the level of a few $\times 10^{-4}$ \citep[e.g.][]{2019arXiv191209481T}.

Detectors contain many effects that can become significant sources of shear systematics if they are not properly modeled.
In contrast to the present generation of optical surveys using silicon-based charge coupled devices (CCDs), {\slshape WFIRST} will rely on near-infrared complementary metal?oxide?semiconductor (CMOS) detector arrays that must use a narrower band gap material. Each detector array consists of a Teledyne H4RG-10 readout integrated circuit hybridized to a mercury cadmium telluride (HgCdTe, 2.5-$\mu$m cutoff) layer that contains the photodiodes, which in turn has an antireflection coating.\footnote{For more detailed information on such devices see \citet{2008SPIE.7021E..0HB} and \citet{2011ASPC..437..383B}. For development of these devices for WFIRST, see \citet{2014SPIE.9154E..2HP} and the overview presentation by B. Rauscher at \url{https://wfirst.gsfc.nasa.gov/science/presentations/vugraphs/160202WFIRSTSDTH4RG10C.pdf}. The ``4'' in the HxRG name indicates that the array has 4k$\times$4k pixels, and the ``10'' indicates that the pixels are on a square grid with 10 $\mu$m spacing.} In order to achieve accurate weak lensing results at the desired precision, we require thorough modeling of any potential biases this physical configuration introduces into measurements of galaxy sizes and ellipticities. The NIR detectors provide some advantages over CCDs -- in particular the non-destructive read capability -- but since they have not been used for weak lensing by large scale structure before, and have historically shown many non-ideal features, {\slshape WFIRST} has put particular emphasis on understanding how NIR detector systematics impact weak lensing \citep[e.g.][]{2013PASP..125.1065S, 2016PASP..128i5001K, 2016PASP..128j4001P}.

In this paper, we are primarily concerned with non-linear behavior arising from the interaction of the brighter-fatter effect (BFE) and inter-pixel capacitance (IPC). Non-linear effects in detectors are of particular concern because we use bright stars to measure the point spread function (PSF), and then use this PSF model to correct the shapes of faint galaxies. The brighter-fatter effect describes the tendency of a brighter image to appear larger due to the deflection of incoming photoelectrons away from ``bright'' pixels that accumulate many electrons (e.g., \citealt{2014JInst...9C3048A}), while the IPC describes electrical cross-talk among neighboring pixels (e.g., \citealt{2004SPIE.5167..204M}). In current optical surveys, the BFE has been shown to introduce a flux dependence into the point-spread function and into correlations among pixels (DES: \citealt{2015JInst..10C5032G}; HSC: \citealt{2018AJ....155..258C}). The effect has also been characterized in prototype devices for upcoming optical surveys including Euclid \citep{2015ExA....39..207N} and LSST \citep{2015JInst..10C5024B, 2017JInst..12C3091L, 2019A&A...629A..36A}. Similarly, characterization of the BFE in infrared devices will be required for {\slshape WFIRST}. Analysis of this kind has been carried out for previous generations of Teledyne detectors (H1R and H2RG) by \citet{2017JInst..12C4009P}, who have also detected the BFE through spot illumination in Euclid prototype devices using multiple up-the-ramp samples \citep{2018PASP..130f5004P}.  

\citet[hereafter Paper I]{2020PASP..132a4501H} presents a formalism for measuring the BFE by correlating flat-field images from H4RG-10 detectors. We note that while this formalism and its extension in this paper is motivated by the need to characterize HgCdTe detectors sensitive in the infrared, it applies generally to CMOS detectors. Flat-field correlations are dominated by the IPC, which includes both linear and signal-dependent components \citep{2016SPIE.9915E..2ID, 2017OptEn..56b4103D, 2018PASP..130g4503D}. However, the non-destructive readout capability of HxRGs allows cross-correlation among different timeslices of a single exposure. Paper I exploits this technique to separate IPC and BFE, while also modeling their interaction with the ``classical'' nonlinearity (CNL), i.e., the non-linearity of converting charge to signal in data numbers (DN). This is a generalization of the traditional photon transfer curve that yields a gain measurement \citep[e.g.,][]{1981SPIE..290...28M, 1985SPIE..570....7J} and the correlation analyses for BFE in CCDs \citep[e.g.,][]{2015A&A...575A..41G} and IPC in infrared arrays \citep[e.g.,][]{2004SPIE.5167..204M}. In addition to the formalism, Paper I presents \textsc{Solid-waffle}, a code that simulates flats and implements algorithms for extracting IPC and BFE parameters from flat-field exposures. The authors conclude that, while the framework is capable of extracting most nonlinearity parameters to high accuracy, a bias of $\sim 12$\% remains for the BFE kernel when the method is applied to simulated data. They posit that this bias arises from high-order nonlinear interactions among IPC, BFE, and CNL whose terms are dropped in the approximations required to solve the  correlation function equation. 

In  Paper II \citep[hereafter Paper II]{2020PASP..132a4502C}, the same authors apply the framework of Paper I to flat-fields measured on WFIRST development detector SCA 18237. They detect a residual correlation signal in these measurements after accounting for CNL, indicating an overall interpixel nonlinearity (IPNL) resulting from a combination of IPC and BFE. With additional tests they establish that BFE dominates over the nonlinear IPC component (NL-IPC) as the main contributor to the IPNL.

In this paper (Paper III), we remedy the bias in the BFE kernel measurement by reworking the formalism in Fourier space\footnote{A Fourier-space formalism for flat-field correlations from CCDs has been presented by \citet{2019A&A...629A..36A}} (see \S\ref{sec:extension}), which preserves the higher-order nonlinearities, and modify the \textsc{Solid-waffle} code accordingly (see \S\ref{sec:implementation}). In \S\ref{sec:characterization}, we test the reworked formalism to measure IPNL parameters for three flight candidate detectors (sensor chip assemblies SCA 20663, SCA 20828, and SCA 20829); to allow direct comparison with Paper II, we also re-apply the framework to the development detector analyzed in Paper II (SCA 18237). These data were acquired at the Detector Characterization Laboratory (DCL) at the NASA Goddard Spaceflight Center.\footnote{\url{https://detectors.gsfc.nasa.gov/DCL/}} We emphasize that while these are flight candidate detectors, the data were acquired under laboratory conditions with a laboratory controller (a Leach controller instead of the ACADIA flight controller, \citealt{2018SPIE10709E..0TL}) and so some parameters will be different in flight. In \S\ref{sec:discussion}, we summarize our findings and discuss areas for future work.

\section{Theoretical background}
\label{sec:theoretical_background}

We begin with an overview of fundamental ideas from Papers I and II which are necessary to describe detector effects such as BFE, IPC, and CNL. In the following subsections we limit ourselves to presenting the principal mathematical formulae; interested readers are encouraged to peruse Section 2 of Papers I and II for more extensive discussions.

\subsection{Infrared detector signals}

Detector output data are provided in units of data numbers (DNs), which are voltages quantized as 16-bit integers. Light hitting the detector causes voltage across the photodiode to decrease, which in turn causes the observed signal $S$ (units: DN) to decrease.\footnote{The voltage on the sensing node increases during observation. Depending on the polarity of the analog-to-digital converter, the signal in DN may increase or decrease; in the {\slshape WFIRST} detector acceptance testing dewar, $S$ decreases.} In a flat field, the relationship between accumulated charge, $Q$, and a drop in signal level is observed to be nonlinear \citep[e.g.][]{2005nicm.rept....2B,2010SPIE.7731E..3CD,2010SPIE.7742E..22H}; this is referred to as classical nonlinearity (CNL), and can arise from many steps in the signal chain (both in the SCA and the controller). This is frequently modeled by a quadratic polynomial
\begin{equation}\label{eq:cnl-quad}
    S_{\textrm{initial}}-S_{\textrm{final}}=\frac{1}{g}(Q-\beta Q^{2}),
\end{equation}
where $g$ is the gain (units: e/DN) and $\beta$ is the quadratic nonlinearity coefficient, though in practice a higher-order polynomial such as equation \ref{eq:cnl-poly} is often needed. Note that we follow standard practice and describe charge in units of ``electrons,'' even though the pixels in the HxRG series detectors actually collect holes (the math is the same).

The pixels do not operate independently. One major source of interaction is interpixel capacitance (IPC), i.e., parasitic capacitance between neighboring pixels \citep[e.g.][]{2004SPIE.5167..204M}. We can modify equation \ref{eq:cnl-quad} to include IPC; for a pixel located at position $i,j$ the signal drop is
\begin{equation}
    S_{\textrm{initial}}(i,j)-S_{\textrm{final}}(i,j)=\frac{1}{g}\sum_{\Delta_{i},\Delta_{j}}[K_{\Delta_{i},\Delta_{j}}+K^I_{\Delta_{i},\Delta_{j}}\overline{Q}]Q_{i-\Delta_{i},j-\Delta_{j}},
\end{equation}
where $\overline{Q}$ is the mean accumulated charge ($It$ in a flat exposure, with current $I$ per pixel given in units of e/s and time $t$ in seconds) and the kernel matrices $K$ and $K^I$  describe linear and nonlinear (hereafter NL-IPC) components of IPC, respectively. The kernel matrix is normalized to satisfy
\begin{equation}
    \sum_{\Delta_{i},\Delta_{j}}K_{\Delta_{i},\Delta_{j}}=1.
\end{equation}
In cases where cross-talk is evenly distributed among the four nearest neighbors $K_{0,0}=1-4\alpha$, $K_{\pm 1,0}=K_{0,\pm 1}=\alpha$, and all others are zero. However we frequently observe asymmetries between horizontal and vertical directions in NIR detectors. Therefore we separately measure $\alpha_{H}=K_{\pm 1,0}$ and $\alpha_{V}=K_{0,\pm 1}$; in the case where these are different we define $\alpha\equiv (\alpha_{H}+\alpha_{V})/2 = K_{\langle 0,1\rangle}$. (We use angular brackets on the subscript of the kernel to denote averages among the positions with the symmetries of a square lattice -- here, $K_{\langle 0,1\rangle}$ is the average of $K_{0,1}$, $K_{1,0}$, $K_{0,-1}$ and $K_{-1,0}$.) We also allow for diagonal IPC $\alpha_{D}=K_{\pm 1,\pm 1}$.

Other effects in the electronics that leak signal from one pixel into another (e.g., settling when we switch from reading one pixel to the next) may also act phenomenologically as if they were IPC, even though the physical cause is not the electrostatics of the pixels. Thus our determinations of ``IPC'' may include contributions from these effects.

We have used a model for NL-IPC in which the coupling coefficient varies with signal level, motivated by both measurement on H2RG detectors and by electrostatics simulations \citep[e.g.][]{2016SPIE.9915E..2ID, 2017OptEn..56b4103D, 2019arXiv191208331D}. However, we will see that in {\slshape WFIRST} detectors, another effect -- the vertical trailing pixel effect (VTPE), which is related to the readout pattern -- is a larger non-linear effect than NL-IPC.

\subsection{The Brighter-fatter Effect}

A pixel $(i,j)$ has an effective area that changes based on the charge of neighboring pixels. This is described by the model of \citet{2014JInst...9C3048A}:
\begin{equation}
    \mathcal{A}_{i,j}=A^{0}_{i,j}\biggl[1+\sum_{\Delta_{i},\Delta_{j}}a_{\Delta_{i},\Delta_{j}}Q(i+\Delta i, j+\Delta j)\biggr],
    \label{eq:Abfe}
\end{equation}
where $A^{0}_{i,j}$ is the original pixel area, the charge $Q(i,j)$ is given in numbers of elementary charges, and $a_{\Delta_{i},\Delta_{j}}$ is the BFE coupling matrix. While $a_{\Delta_{i},\Delta_{j}}$ is formally dimensionless, it will normally be given in units of $10^{-6} e^{-1}$, ppm/e, or, equivalently, $\%/10^{4}e$. This is a convenient choice because $10^{4}e$ is the typical integrated signal level in the central pixel of a standard PSF star used for \textit{WFIRST}, so a measurement in these units maps to a percent-level effect on a star. In this work we will assume the BFE obeys discrete translational invariance, as in Eq.~(\ref{eq:Abfe}), at least over a large enough group of pixels to do statistics.

\subsection{Correlation Functions}

Infrared array flats typically allow multiple samples to be obtained up the ramp, permitting us to measure correlations both between different pixels and between different frames. It is these temporal correlations which provide the key to disentangling the BFE and NL-IPC (see Papers I and II). Denoting different frames $abcd$, we define the correlation function
\begin{equation}
    C_{abcd}(\Delta i,\Delta j)=\textrm{Cov}[S_{a}(i,j)-S_{b}(i,j),S_{c}(i+\Delta i,j+\Delta j)-S_{d}(i+\Delta i,j+\Delta j)],
\end{equation}
which satisfies four properties outlined in Section 2.3 of Paper I. For this work we assume $a<b$ and $c<d$ but make no further assumptions about time ordering; the intervals may be the same, may overlap, or may be disjoint.

\begin{table}[h]
    \centering
    \begin{tabular}{|c|c|l|}
    \hline\hline
    Quantity      &Units &Description  \\
    \hline
    $Q$   &ke   &Charge, current multiplied by time. \\
    \hline
   $g$ &e/DN &Gain, corrected for IPC and classical non-linearity unless\\ & &specified (e.g. subscript `raw').     \\
   \hline
   $K$ & &IPC kernel matrix, with $K_{0,0} = 1-4\alpha$, $K_{0,\pm 1} = K_{\pm 1,0} = \alpha$. \\
   \hline
   $\alpha$ &\% &Specifies the IPC kernel, average of horizontal (subscript `H')\\&& and vertical (subscript `V') components. Diagonal component\\ & &denoted with subscript `D'.\\
   \hline
  $K^I$ & &Signal level-dependent NL-IPC kernel matrix ($3\times3$). \\&& Equivalent to $K'$ in Paper I. \\
  \hline
   $\beta$ &ppm/e & Leading order classical non-linearity coefficient. \\
   \hline
    $a_{\Delta x_1, \Delta x_2}$ & ppm/e &BFE kernel coefficients defined in terms of shifts from the \\&&central pixel ($\Delta x_1 = \Delta x_2=0$). \\
    \hline
    $\suma$ & ppm/e &Sum of $a_{\Delta x_1,\Delta x_2}$ over $\Delta x_1$,$\Delta x_2$. \\
    \hline
    $[K^{2}a+KK^I]_{\Delta x_1, \Delta x_2}$ &ppm/e &Inter-pixel non-linearities (IPNL) including linear IPC, \\&&non-linear IPC, and BFE. Terms inside brackets are \\&& convolved. \\
    \hline
    \end{tabular}
    \caption{Summary of detector parameters.}
    \label{tab:param_summary}
\end{table}

\section{Fourier-domain treatment of correlations}
\label{sec:extension}

Paper I provides analytic equations for $\mathrm{Cov}[Q(i,j;t),Q(i',j';t_1)]$, the charge covariance matrix, by (i) propagating the change in pixel boundaries produced by the BFE at time $t$ to the one- and two-point functions of pixel charge at time $t+\delta t$ and (ii) approximately solving the resulting differential equations by Taylor expanding to 2nd order, i.e. including terms of order $\alpha^2$, $\alpha\beta$, and $\alpha a$ (see section 3.5 of Paper I). The resulting covariance may then be convolved appropriately with $K$ or otherwise manipulated to account for IPC and nonlinear effects (see e.g. equation 48 of Paper I). While this method results in elegant and intuitive equations for the correlation function, it ignores higher-order terms that are not in fact negligible. 

In what follows, we develop an alternative, fully analytic solution for the nonlinear correlation function by performing the calculations in Fourier space. This method aims to rectify the bias introduced by approximating away high-order effects in the configuration space solution. The calculations presented below follow a similar sequence to those presented in sections 3.5-3.7 of Paper I, beginning with a BFE-only charge covariance and extending this result to include other effects. As such, we refer to equations from Paper I in the format ``equation I/[number]''.

\subsection{Definitions and initial transformations}

We use the following convention for the Fourier transform and its inverse, respectively:
\begin{equation}\label{eqn:fourier}
    \widetilde{F}(k_1,k_2)=\sum_{x_1=0}^{N-1}\sum_{ x_2=0}^{N-1}F(x_1,x_2)\exp\left[\frac{-2\pi i(k_{1} x_1+k_{2} x_{j})}{N}\right]
\end{equation}
\begin{centering}
and
\begin{equation}
    {F}(x_1,x_2)=\frac{1}{N^2}\sum_{k_1=0}^{N-1}\sum_{ k_2=0}^{N-1}\widetilde{F}(k_1,k_2)\exp\left[\frac{2\pi i(k_{1} x_1+k_{2} x_{2})}{N}\right].
\end{equation}
\end{centering}

\noindent Here, $(2\pi)(k_1,k_2)/N$ is a wave vector in radians per pixel, where the detector is assumed to comprise an $N\times N$ array of pixels; similarly, $(u_i,u_j)=(k_1,k_2)/N$ is a wave vector in cycles per pixel. Note that while Paper I uses $(i,j)$ to denote position on the detector, we use $(x_1,x_2)$, in order to enable the reader to distinguish between equations in Fourier space and equations in configuration space. Throughout, we assume periodic boundary conditions, and we will often notate a double summation from from 0 to $N-1$ as $\sum_{x_1,x_2}$ or similar.

We construe the BFE as a change in the effective area of a pixel. This area defect $W$ at time $t$ is given by equation I/31, 
\begin{equation}
    W(x_{1},x_{2},t) = 1 + \sum_{\Delta x_{1},\Delta x_{2}}a_{\Delta x_1,\Delta x_2}Q(
    x_1+\Delta x_1,x_2+\Delta x_2,t),
\end{equation}
where $a$ is the BFE kernel and $Q$ is charge. We apply the Fourier transform to yield $\widetilde{W}$, 
\begin{equation}
\begin{aligned}
    \widetilde W(k_1,k_2,t)&= N^2\delta_{k_1,0}\delta_{k_2,0}+\sum_{x_1,x_2}\left[\sum_{\Delta x_1,\Delta x_2}a_{\Delta x_1, \Delta x_2}Q(x_1+\Delta x_1, x_2+\Delta x_2,t)\right] \\
    &\times\exp\left(-\frac{2\pi i}{N}(k_1 x_1+k_2 x_2)\right)
\end{aligned}
\end{equation}
By defining a sign-flipped BFE kernel $a^\mathrm{flipped}_{\Delta x_1,\Delta x_2} \equiv a_{-\Delta x_1,-\Delta x_2}$, we can rewrite the terms inside the brackets above as a convolution, i.e.
\begin{equation}
    \sum_{-\Delta x_1,-\Delta x_2}a^\mathrm{flipped}_{\Delta x_1, \Delta x_2}Q(x_1-\Delta x_1, x_2-\Delta x_2,t) = (a^\mathrm{flipped}*Q)(x_1,x_2,t).
\end{equation}
We note that $\widetilde{a}^*(k_1,k_2)=\widetilde{a^\mathrm{flipped}}(k_1,k_2)$, where $\widetilde{a}^*$ is the complex conjugate of $\widetilde{a}$. Then, by the convolution theorem,
\begin{equation}
    \widetilde{W}(k_1,k_2,t) = N^2\delta_{k_1,0}\delta_{k_2,0} + \widetilde{a}^*(k_1,k_2)\widetilde{Q}(k_1,k_2,t)
\end{equation}

\subsection{Brighter-fatter effect}
A summary of the notation used in the following calculations is shown in table \ref{tab:notation_summary}. These definitions are noted in the text below where applicable.

\begin{table}[h]
    \centering
    \renewcommand{\arraystretch}{1.2}
    \begin{tabular}{|c|c|l|}
    \hline\hline
    Notation      &Definition &Notes  \\
    \hline
    $\widetilde{F}$ & Fourier transform of $F$ & Equation \ref{eqn:fourier}\\
    \hline
    $\widetilde{Q}(t)$  & $\widetilde{Q}(k_1,k_2,t)$ & Where $t$ indicates a timeframe $abcd$ \\ && we subscript accordingly, e.g. $\widetilde{Q}_a\equiv\widetilde{Q}(t_a)$   \\
    \hline
    $\widetilde{Q}'(t)$  & $\widetilde{Q}(k_1',k_2',t)$ & note above applies  \\
    \hline
    $\widetilde{a}^*$  & $\widetilde{a}^*(k_1,k_2)$ & Similar for $K$, $K^I$  \\
    \hline
    $\widetilde{a}^*{'}$  & $\widetilde{a}^*(k_1',k_2')$ &  note above applies \\
    \hline
    $\widetilde{a}^*{^+}$  & $\widetilde{a}^*(k_1+k_1',k_2+k_2')$ &   \\
    \hline
    $\delta_{k,k',0}$  & $\delta_{k_1,0}\delta_{k_1',0}\delta_{k_2,0}\delta_{k_2',0}$ & Equation \ref{eqn:deltas} \\
    \hline
    $\delta_{k+k',0}$  & $\delta_{k_1+k_1',0}\delta_{k_2+k_2',0}$ & Equation \ref{eqn:deltas}  \\
    \hline
    $S^{n,\mathrm{CNL}}_a$ & $S^\mathrm{CNL}_a$ expressed as a & Equation \ref{eq:cnl-poly}\\& polynomial of degree $n$ & \\
    \hline
    
    \end{tabular}
    \caption{Summary of notation used in the calculations in \S\ref{sec:extension}.}
    \label{tab:notation_summary}
\end{table}

\subsubsection{Equal-time correlation function}
\label{sssec:equaltime}
Given the state of the system $\widetilde{Q}(k_1,k_2,t)$ at time $t$, we can write the mean charge in mode $(k_1,k_2)$ at time $t+\delta t$ (following equations I/32 and I/33) as the initial charge plus a residual due to the BFE:
\begin{equation}\label{eqn:firstmoment}
 \begin{aligned}
    \langle\widetilde{Q}(k_1,k_2,t+\delta t)\rangle|_t =  \widetilde Q(k_1,k_2,t) + \Delta \widetilde Q (k_1,k_2,t) = \widetilde Q(k_1,k_2,t) + I\widetilde W(k_1,k_2,t)\delta t.
 \end{aligned}
\end{equation}
We assume that $I\delta t<<1$; that is, we are operating in the single-electron limit, and $IW(x_1,x_2,t)\delta t$ corresponds to the probability of pixel $(x_1,x_2)$ collecting a that electron during time interval $\delta t$. The expectation value of equation \ref{eqn:firstmoment} is
\begin{equation}
\begin{split}
\langle\widetilde  Q(k_1,k_2,t+\delta t)\rangle= \langle\widetilde Q(k_1,k_2,t)\rangle &+ IN^2\delta_{k_1,0}\delta_{k_2,0}\delta t + I\widetilde{a}^*(k_1,k_2)\langle\widetilde Q(k_1,k_2,t)\rangle \delta t.
\end{split}
\end{equation}
Taking the limit as $\delta t$ goes to 0, we may then write
\begin{equation}
\begin{split}
\frac{d}{dt}\langle\widetilde  Q(k_1,k_2,t)\rangle=  IN^2\delta_{k_1,0}\delta_{k_2,0}
+ I\widetilde{a}^*(k_1,k_2)\langle\widetilde Q(k_1,k_2,t)\rangle,
\end{split}
\end{equation}
which is a first order ODE with the solution
\begin{equation}\label{eqn:fm_solution}
\langle \widetilde Q(k_1,k_2,t) \rangle = \frac{N^2}{\widetilde{a}^*(k_1,k_2)} \left(e^{I\widetilde{a}^*(k_1,k_2)t}-1\right)\delta_{k_1,0}\delta_{k_2,0}
\end{equation}
for $\langle\widetilde Q(k_1,k_2,t)\rangle = 0$ at $t=0$.

Similarly, we may write the two-point function for charge in modes $(k_1,k_2)$ and $(k_1',k_2')$ at time $t+\delta t$, given the states $\widetilde{Q}(k_1,k_2,t)$ and $\widetilde{Q}(k_1',k_2',t)$ at time $t$:
\begin{equation}\label{eqn:secondmoment}
 \begin{aligned}
    \langle\widetilde{Q}(k_1,k_2,&t+\delta t)\widetilde{Q}(k_1',k_2',t+\delta t)\rangle|_t =  
    \widetilde{Q}(k_1,k_2,t)\widetilde{Q}(k_1',k_2',t)
    + I\widetilde W(k_1,k_2,t)\widetilde{Q}(k_1',k_2',t)\delta t \\
    &+  I\widetilde W(k_1',k_2',t)\widetilde{Q}(k_1,k_2,t)\delta t
    + \mathrm{Cov}[\Delta\widetilde Q (k_1,k_2,t),\Delta\widetilde Q (k_1',k_2',t)].
 \end{aligned}
\end{equation}
The configuration-space covariance term $\mathrm{Cov}[\Delta Q (x_1,x_2,t),\Delta Q (x_1',x_2',t)]$ is only nonzero when $x_1=x_2$, since otherwise two electrons would be required to increment both $Q(x_1,x_2,t)$ and $Q(x_1',x_2',t)$. Therefore, we may write the covariance term in Fourier space as
\begin{equation}
 \begin{split}
    \mathrm{Cov}[\Delta\widetilde Q (k_1,k_2,t),\Delta\widetilde Q (k_1',k_2',t)] &= \sum_{x_1,x_2}IW(x_1,x_2,t)\delta t 
    \times \exp\left(\frac{-2\pi i}{N}\big((k_1+k_1')x_1+(k_2+k_2')x_2\big)\right) \\
    &=I\widetilde W(k_1+k_1',k_2+k_2',t)\delta t.
 \end{split}
\end{equation}
With this result, the expectation value of equation \ref{eqn:secondmoment} (taking the limit $\delta t\rightarrow 0$ as before and rearranging) is
\begin{equation}\label{eqn:twopoint_diffeq}
 \begin{aligned}
    \frac{d}{dt}\langle\widetilde{Q}(k_1,k_2,t)\widetilde{Q}(k_1',k_2',t)\rangle = &\;I\left(\widetilde{a}^*(k_1,k_2)+\widetilde {a}^*(k_1',k_2')\right)\langle \widetilde Q(k_1,k_2,t)\widetilde Q(k_1',k_2',t) \rangle  \\
    &+IN^2\delta_{k_1',0}\delta_{k_2',0}\langle\widetilde Q(k_1,k_2,t)\rangle  +IN^2\delta_{k_1,0}\delta_{k_2,0}\langle\widetilde Q(k_1',k_2',t)\rangle  \\
    &+I\widetilde{a}^*(k_1+k_1',k_2+k_2')\langle\widetilde Q(k_1+k_1',k_2+k_2',t)\rangle  \\
    &+IN^2\delta_{k_1+k_1',0}\delta_{k_2+k_2',0}
 \end{aligned}
\end{equation}
From here on, we refer to $\widetilde{a}^{*}(k_1',k_2')$ as $\widetilde{a}^{*\prime}$ and $\widetilde{a}^{*}(k_1+k_1',k_2+k_2')$ as $\widetilde{a}^{*+}$. Similarly we write $\widetilde{Q}(k_1',k_2',t)$ as $\widetilde{Q}'(t)$. We make the definitions
\begin{equation}\label{eqn:deltas}
\delta_{k,k',0}\equiv\delta_{k_1,0}\delta_{k_2,0}\delta_{k_1',0}\delta_{k_2',0} ~~~{\rm and}~~~
\delta_{k+k',0}\equiv\delta_{k_1+k_1',0}\delta_{k_2+k_2',0}.
\end{equation}
Then substituting equation \ref{eqn:fm_solution} into equation \ref{eqn:twopoint_diffeq} yields another first-order differential equation, 
\begin{equation}\label{eqn:dqqdt}
 \begin{aligned}
    \frac{d}{dt}\langle\widetilde{Q}(t)\widetilde{Q}'(t)\rangle &= \langle \widetilde Q(t)\widetilde Q'(t) \rangle \times I\left(\widetilde{a}^*+\widetilde{a}^{*\prime}\right) + \\
    &IN^4\delta_{k,k',0}\left(\frac{e^{I\widetilde{a}^{*}t}-1}{\widetilde{a}^{*}} + \frac{e^{I\widetilde{a}^{*\prime}t}-1}{\widetilde{a}^{*\prime}}\right) +
    IN^2e^{I\widetilde{a}^{*+} t}\delta_{k+k',0}.
 \end{aligned}
\end{equation}
For the initial condition $\langle\widetilde{Q}(t=0)\widetilde{Q}'(t=0)\rangle = 0$, we may write the exact solution for the equal-time correlation function:
\begin{equation}\label{eqn:equaltime}
 \begin{split}
    \langle\widetilde{Q}(t)\widetilde{Q}'(t)\rangle = \;\frac{N^4\delta_{k,k',0}}{\widetilde{a}^*\widetilde{a}^{*\prime}}\Big(e^{I\widetilde{a}^*t}-1\Big)\left(e^{I\widetilde{a}^{*\prime}t} - 1\right) +\frac{N^2\delta_{k+k',0}}{\widetilde{a}^*+\widetilde{a}^{*\prime}-\widetilde{a}^{*+}}\Big(e^{I\left(\widetilde{a}^*+\widetilde{a}^{*\prime}\right)t}-e^{I\widetilde{a}^{*+}t}\Big).
 \end{split}
\end{equation}

\subsubsection{Unequal-time correlation function}
\label{sssec:unequaltime} 

In addition to the equal-time correlation function, we need to solve for the correlation between mode $(k_1,k_2)$ at time $t$ and mode $(k_1',k_2')$ at time $t_1$. To obtain this equation, we first multiply equation \ref{eqn:firstmoment} by $\widetilde{Q}(k_1',k_2',t_1)$. The propagation of the BFE is a Markovian process, i.e. only dependent on the state of the system at one previous time $t$ rather than multiple previous times; therefore, we may write
\begin{equation}\label{eqn:firstmoment_unequal}
 \begin{aligned}
    \langle\widetilde{Q}(k_1,k_2,t+\delta t)\widetilde{Q}(k_1',k_2',t_1)\rangle|_{t,t_1} =  \widetilde Q(k_1,k_2,t)\widetilde{Q}(k_1',k_2',t_1) + I\widetilde W(k_1,k_2,t)\widetilde{Q}(k_1',k_2',t_1)\delta t.
 \end{aligned}
\end{equation}
Taking the expectation value and taking the limit as $\delta t \rightarrow 0$ produces an ODE:
\begin{equation}
\begin{split}
    \frac{d}{dt}\langle\widetilde{Q}(t) \widetilde{Q}'(t_1)\rangle &= I\langle \widetilde{Q}'(t_1)\widetilde{W}(t)\rangle =IN^2\delta_{k_1,0}\delta_{k_2,0}\langle\widetilde{Q}'(t_1)\rangle + I\widetilde{a}^*\langle\widetilde{Q}'(t_1)\widetilde{Q}(t)\rangle.
\end{split}
\end{equation}
The solution is given by
\begin{equation} \label{eqn:sm_solution}
    \langle\widetilde{Q}(t)\widetilde{Q}'(t_1)\rangle = \frac{N^4\delta_{k,k',0}}{\widetilde{a}^*\widetilde{a}^{*\prime}}\Big(e^{I\widetilde{a}^{*\prime}t_1}-1\Big)\Big(e^{I\widetilde{a}^*t}-1\Big)+ \frac{N^2\delta_{k+k',0}}{\widetilde{a}^*+\widetilde{a}^{*\prime}-\widetilde{a}^{*+}}\;e^{I\widetilde{a}^*(t-t_1)}\left(e^{I\left(\widetilde{a}^*+\widetilde{a}^{*\prime}\right)t_1}-e^{I\widetilde{a}^{*+}t_1}\right),
\end{equation}
where we have substituted for $\langle\widetilde{Q}'(t_1)\rangle$ using equation \ref{eqn:fm_solution} and applied the condition that the equal-time correlation function (equation \ref{eqn:equaltime}) holds at $t=t_1$. 

\subsubsection{Signal correlation}
\label{sssec:fullcorrelation}

We define the Fourier-space correlation function of the signal (i.e. the power spectrum) across times $(t_a,t_b,t_c,t_d)$ to be
\begin{equation}
    \widetilde{C}_{abcd}(k_1'-k_1,k_2'-k_2) = \mathrm{Cov}\left[\widetilde{S}_a(k_1,k_2)-\widetilde{S}_b(k_1,k_2),\widetilde{S}_c(k_1',k_2')-\widetilde{S}_d(k_1',k_2')\right],
\end{equation}
where $\widetilde{S}_0 - \widetilde{S}_a = \widetilde{Q}_a/g^2$, in the absence of IPC and any terms beyond linear order. The first term of equation \ref{eqn:sm_solution} is equivalent to $\langle\widetilde{Q}(t)\rangle\times\langle\widetilde{Q}'(t_1)\rangle$, and so the charge covariance is given by the second term, i.e.
\begin{equation}\label{eqn:covariance}
    \mathrm{Cov}\left[\widetilde{Q}(t),\widetilde{Q}'(t_1)\right] = \frac{N^2\delta_{k+k',0}}{\widetilde{a}^*+\widetilde{a}^{*\prime}-\widetilde{a}^{*+}}\;e^{I\widetilde{a}^*(t-t_1)}\left(e^{I\left(\widetilde{a}^*+\widetilde{a}^{*\prime}\right)t_1}-e^{I\widetilde{a}^{*+}t_1}\right).
\end{equation}

Then, considering only the linear response of the detector and defining $\widetilde{Q}_a\equiv \widetilde{Q}(t_a)$,
\begin{equation} \label{eqn:c_bfe}
\begin{split}
    \widetilde{C}^\mathrm{BFE}_{abcd}&(k_1'-k_1,k_2'-k_2)=\frac{1}{g^2}\bigg\{\mathrm{Cov}\left[\widetilde{Q}_a\widetilde{Q}'_c\right]-\mathrm{Cov}\left[\widetilde{Q}_a\widetilde{Q}'_d\right]-\mathrm{Cov}\left[\widetilde{Q}_b\widetilde{Q}'_c\right]+\mathrm{Cov}\left[\widetilde{Q}_b\widetilde{Q}'_d\right]\bigg\} \\
    &=\frac{N^2\delta_{k+k',0}}{g^2\left(\widetilde{a}^*+\widetilde{a}^{*\prime}-\widetilde{a}^{*+}\right)}\Bigg[e^{I\widetilde{a}^*\big(t_{\mathrm{max}(a,c)}-t_{\mathrm{min}(a,c)}\big)}\left(e^{I\left(\widetilde{a}^*+\widetilde{a}^{*\prime}\right)t_{\mathrm{min}(a,c)}}-e^{I\widetilde{a}^{*+}t_{\mathrm{min}(a,c)}}\right) \\
    &\hspace{77pt}-e^{I\widetilde{a}^*\big(t_{\mathrm{max}(a,d)}-t_{\mathrm{min}(a,d)}\big)}\left(e^{I\left(\widetilde{a}^*+\widetilde{a}^{*\prime}\right)t_{\mathrm{min}(a,d)}}-e^{I\widetilde{a}^{*+}t_{\mathrm{min}(a,d)}}\right) \\
    &\hspace{77pt}-e^{I\widetilde{a}^*\big(t_{\mathrm{max}(b,c)}-t_{\mathrm{min}(b,c)}\big)}\left(e^{I\left(\widetilde{a}^*+\widetilde{a}^{*\prime}\right)t_{\mathrm{min}(b,c)}}-e^{I\widetilde{a}^{*+}t_{\mathrm{min}(b,c)}}\right) \\
    &\hspace{77pt}+e^{I\widetilde{a}^*\big(t_{\mathrm{max}(b,d)}-t_{\mathrm{min}(b,d)}\big)}\left(e^{I\left(\widetilde{a}^*+\widetilde{a}^{*\prime}\right)t_{\mathrm{min}(b,d)}}-e^{I\widetilde{a}^{*+}t_{\mathrm{min}(b,d)}}\right)\Bigg].
\end{split}
\end{equation}

\subsection{Interaction of BFE with other effects}

\subsubsection{IPC}

For a perfect, linear detector, the Fourier transform of the signal corresponding to a voltage decrease between time $t=0$ and time $t=t_a$ is given by
\begin{equation}
    \widetilde{S}_a^\mathrm{perfect}(k_1,k_2) = \frac{1}{g}\widetilde{Q}(k_1,k_2,t_a).
\end{equation}
This direct linear relationship applies in the presence of BFE, since BFE affects the destination of each incoming electron before it is stored in a pixel. On the other hand, the linear component of IPC acts once the charge is stored. This cross-talk effect is expressed as a convolution in configuration space (see equation I/8), and we can therefore write
\begin{equation}
    \widetilde{S}^\mathrm{IPC}_a(k_1,k_2) = \frac{1}{g}\widetilde{K}(k_1,k_2)\widetilde{Q}(k_1,k_2,t_a),
\end{equation}
where $\widetilde{K}$ is the Fourier transform of the IPC convolution kernel.

\subsubsection{NL-IPC}

The nonlinear IPC effect in configuration space may similarly be written as the convolution of a non-linear kernel with charge (see equation I/10), which we express as a product in Fourier space:
\begin{equation}\label{eqn:s_ipc,nlipc}
    \widetilde{S}^\mathrm{IPC+NLIPC}_a = \frac{1}{g}\widetilde{Q}(k_1,k_2,t_a)\left(\widetilde{K}(k_1,k_2)+\widetilde{K}^I(k_1,k_2)\overline{Q}_a\right).
\end{equation}
Note that in order to avoid confusion we have written $\widetilde{K}^I$ instead of $\widetilde{K}'$, used in Paper I, since we are already using primed notation to refer different locations (in configuration space) or different modes (in Fourier space) on the detector. The correlation function is then
\begin{equation}
\begin{split}
    \widetilde{C}_{abcd}^{\textrm{BFE+IPC+NLIPC}} =& \frac{1}{g^2}\Big\{ (\widetilde{K}+\widetilde{K}^I\overline{Q}_a)(\widetilde{K}'+\widetilde{K}^I{'}\overline{Q}_c)\mathrm{Cov} [\widetilde{Q}_a,\widetilde{Q}_c'] \\
    &-(\widetilde{K}+\widetilde{K}^I\overline{Q}_a)(\widetilde{K}'+\widetilde{K}^I{'}\overline{Q}_d)\mathrm{Cov} [\widetilde{Q}_a,\widetilde{Q}_d'] \\
    &-(\widetilde{K}+\widetilde{K}^I\overline{Q}_b)(\widetilde{K}'+\widetilde{K}^I{'}\overline{Q}_c)\mathrm{Cov} [\widetilde{Q}_b,\widetilde{Q}_c'] \\
    &+(\widetilde{K}+\widetilde{K}^I\overline{Q}_b)(\widetilde{K}'+\widetilde{K}^I{'}\overline{Q}_d)\mathrm{Cov} [\widetilde{Q}_b,\widetilde{Q}_d']
    \Big\},
\end{split}
\end{equation}
where $\mathrm{Cov}[\widetilde{Q}_a,\widetilde{Q}_c']$ and similar terms are the BFE-inclusive covariance given in equation \ref{eqn:covariance}.

\subsubsection{CNL}

Classical non-linearity (CNL) applies to the conversion from stored charge to voltage. Written without IPC effects, the real-space signal is given by
\begin{equation}
    S_a^{n,\mathrm{CNL}}(x_1,x_2) = \frac{1}{g}\left[Q(x_1,x_2,t_a)-\beta_2 Q(x_1,x_2,t_a)^2-...-\beta_n Q(x_1,x_2,t_a)^n\right].
\label{eq:cnl-poly}
\end{equation}
Orders beyond $n=2$ are often dropped (in which case $\beta\equiv\beta_2$), but here we choose a fully general polynomial to allow characterization of CNL to arbitrary order. CNL does not lend itself to an exact analytic solution in Fourier space, since beyond first order $\widetilde{S}_a^\mathrm{CNL}$ includes the autoconvolution of charge. However, since we are working with small fluctuations around the mean charge, we can make the Taylor approximation 
\begin{equation}
S_a^{n,\mathrm{CNL}} \approx \frac{1}{g}\left[\bigg(\overline{Q}_a-\sum_{\nu=2}^n\beta_\nu\overline{Q}_a^\nu\bigg)+\left(Q_a-\overline{Q}_a\right)\bigg(1-\sum_{\nu=2}^n \nu\beta_\nu\overline{Q}_a^{\nu-1}\bigg)\right],
\end{equation}
using the notation $Q_a \equiv Q(x_1,x_2,t_a)$. Then, rearranging and applying the Fourier transform,
\begin{equation}\label{eqn:s_cnl}
\begin{split}
    \widetilde{S}_a^{n,\mathrm{CNL}}=\frac{1}{g}\left[\widetilde{Q}_a\bigg(1-\sum_{\nu=2}^n\nu\beta_\nu\overline{Q}_a^{\nu-1}\bigg)+N^2\delta_{k_1,0}\delta_{k_2,0}\sum_{\nu=2}^n\beta_\nu(\nu-1)\overline{Q}_a^\nu\right].
\end{split}
\end{equation}
Extrapolating from equations \ref{eqn:s_ipc,nlipc} and \ref{eqn:s_cnl}, we can write the signal accounting for IPC, NLIPC, and CNL:
\begin{equation}
\begin{split}
    \widetilde{S}_{a}^{n,\mathrm{IPC+}}&^{\mathrm{NLIPC+CNL}} = \\ &\frac{1}{g}\left[\widetilde{Q}_a\left(\widetilde{K}+\widetilde{K}^{I}\overline{Q}_{a}\right)\bigg(1-\sum_{\nu=2}^n\nu\beta_\nu\overline{Q}_a^{\nu-1}\bigg)+N^2\delta_{k_1,0}\delta_{k_2,0}\sum_{\nu=2}^n\beta_\nu(\nu-1)\overline{Q}_a^\nu\right].
\end{split}
\end{equation}
We are now in a position to write down the full power spectrum, $\widetilde{C}_{abcd}^{\textrm{BFE}+\textrm{IPC}+\textrm{NLIPC}+\textrm{CNL}} \equiv \widetilde{C}_{abcd}^{\textrm{full}}$:
\begin{equation}\label{eqn:fullcorr}
\begin{aligned}
\widetilde{C}_{abcd}^{\textrm{full}}
&= \frac{1}{g^{2}}\left[\bigg(1-\sum_{\nu=2}^n\nu\beta_\nu\overline{Q}_a^{\nu-1}\bigg)\bigg(1-\sum_{\nu=2}^n\nu\beta_\nu\overline{Q}_c^{\nu-1}\bigg)\left(\widetilde{K}+\widetilde{K}^{I}\overline{Q}_{a}\right)\left(\widetilde{K}'+\widetilde{K}^{I\prime}\overline{Q}_{c}\right)\textrm{Cov}\left(\widetilde{Q}_{a},\widetilde{Q}'_{c}\right)\right. \\ 
&\hspace{20pt}-\bigg(1-\sum_{\nu=2}^n\nu\beta_\nu\overline{Q}_a^{\nu-1}\bigg)\bigg(1-\sum_{\nu=2}^n\nu\beta_\nu\overline{Q}_d^{\nu-1}\bigg)\left(\widetilde{K}+\widetilde{K}^{I}\overline{Q}_{a}\right)\left(\widetilde{K}'+\widetilde{K}^{I\prime}\overline{Q}_{d}\right)\textrm{Cov}\left(\widetilde{Q}_{a},\widetilde{Q}'_{d}\right) \\ 
&\hspace{20pt}-\bigg(1-\sum_{\nu=2}^n\nu\beta_\nu\overline{Q}_b^{\nu-1}\bigg)\bigg(1-\sum_{\nu=2}^n\nu\beta_\nu\overline{Q}_c^{\nu-1}\bigg)\left(\widetilde{K}+\widetilde{K}^{I}\overline{Q}_{b}\right)\left(\widetilde{K}'+\widetilde{K}^{I\prime}\overline{Q}_{c}\right)\textrm{Cov}\left(\widetilde{Q}_{b},\widetilde{Q}'_{c}\right) \\ 
&\hspace{20pt}+\left.\bigg(1-\sum_{\nu=2}^n\nu\beta_\nu\overline{Q}_b^{\nu-1}\bigg)\bigg(1-\sum_{\nu=2}^n\nu\beta_\nu\overline{Q}_d^{\nu-1}\bigg)\left(\widetilde{K}+\widetilde{K}^{I}\overline{Q}_{b}\right)\left(\widetilde{K}'+\widetilde{K}^{I\prime}\overline{Q}_{d}\right)\textrm{Cov}\left(\widetilde{Q}_{b},\widetilde{Q}'_{d}\right)\right],
\end{aligned}
\end{equation}
where the covariances are of the form shown in equation \ref{eqn:covariance}. Taking the inverse Fourier transform of equation \ref{eqn:fullcorr} yields the full correlation function $C^\mathrm{full}_{abcd}$, as desired for detector characterization.

\section{Incorporation in detector characterization code}
\label{sec:implementation}

We now proceed to incorporate these results in the {\sc Solid-waffle} code that we are using to characterize WFIRST detectors. We first describe the update to the determination of the classical non-linearity curve in {\sc Solid-waffle} (which now allows for a higher-order than quadratic polynomial). We then describe the code implementation of the full correlation function (\S\ref{ss:code}), and then tests on simulated data (\S\ref{ss:sims}). We recall that {\sc Solid-waffle} takes a set of $N_{\rm F}$ flats and $N_{\rm F}$ darks, and returns parameter measurements in ``super-pixels'' of configurable size since the correlation function measurement is statistical and normally one has to average many pixels to obtain high-S/N results (default: $32\times 32$ super-pixels, each containing $128\times 128$ pixels). Paper I contains implementation details that will not be repeated here.

\subsection{The classical non-linearity curve}
\label{ss:cnl-curve}

The original version of {\sc Solid-waffle} (Paper I) worked with a quadratic approximation to the classical non-linearity -- i.e., we kept only the $\beta_2$ term in Eq.~(\ref{eq:cnl-poly}). When we want to go to large signal levels, the deviation from a quadratic polynomial becomes important, and {\sc Solid-waffle} must determine the higher-order coefficients. We do this by the standard method of multiple samples up the ramp.

We do an unweighted fit of the polynomial to the time slices, ${\rm Median}(S_{1a}) = \sum_{j=0}^p c_j t_a^j$, where $p$ is the order of the polynomial. This fit is performed after subtraction of the left and right reference pixels.\footnote{The reference pixels are non-light-sensitive pixels around the top, bottom, left, and right sides of the detector array that can be used to monitor electronic drifts. Here we use the left and right pixels to remove the horizontal banding seen in, e.g., Fig.~\ref{fig:dfs}, which we found in Paper II to be an effective way to correct pixel medians. It is possible that this procedure may need to be revisited in the flight setup.} Then we normalize the slo to subtrape to $1$ by defining $\bar c_j = c_j/c_1^j$. By construction $\bar c_1=1$. The coefficients $\bar c_j$ have units of DN$^{1-j}$, and are reported in the output files. The non-linearity coefficients in electron units are related to these via
\begin{equation}
\beta_j = -\frac{\bar c_j}{g^{j-1}};
\label{eq:bcg}
\end{equation}
however the gain $g$ is not yet known.

\subsection{Implementation of the full correlation function}
\label{ss:code}

\begin{figure}
    \centering
    \includegraphics[width=6.5in]{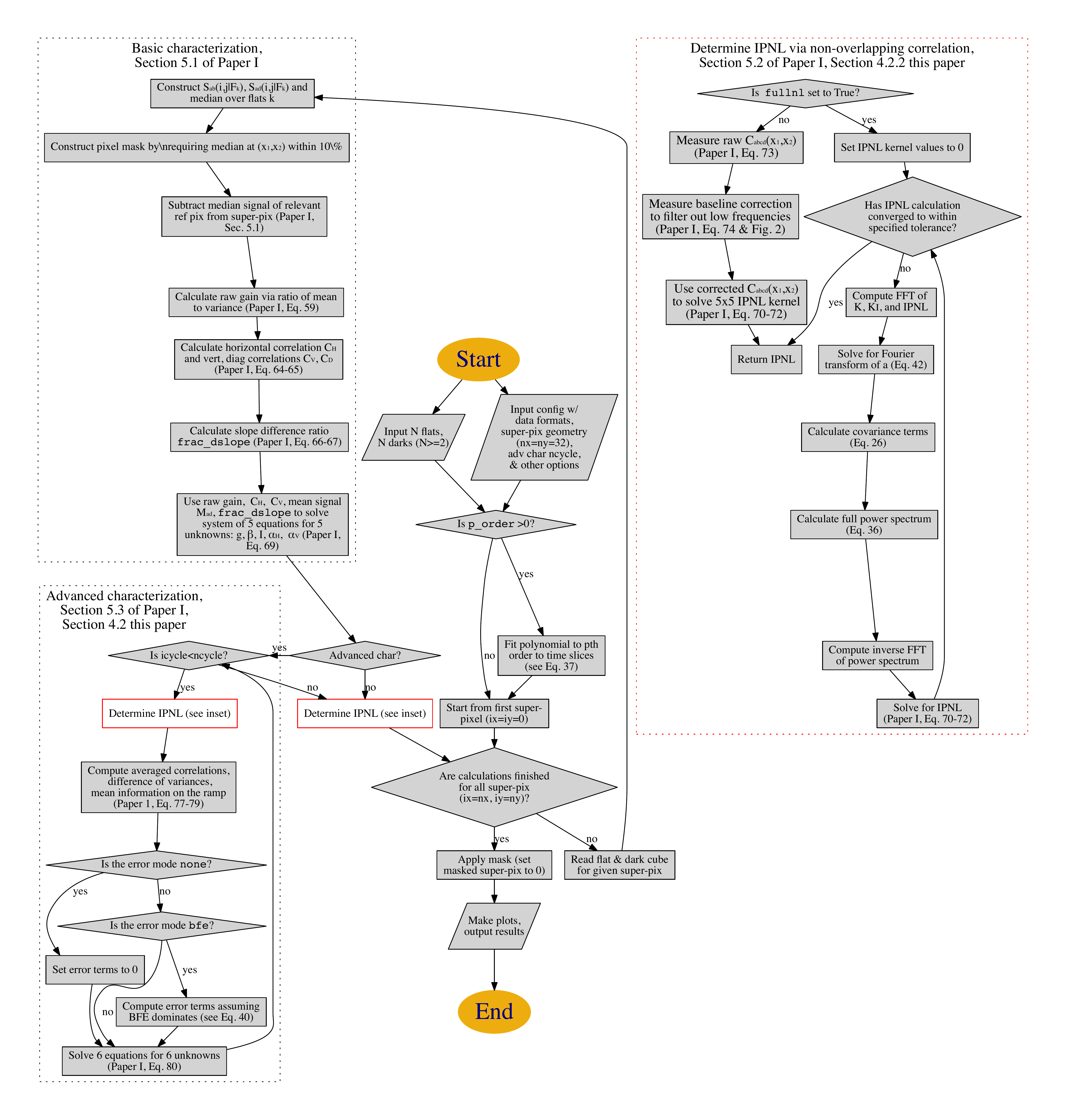}
    \caption{The flow chart for the characterization code.  The main updates relative to Papers I and II are to the IPNL determination (inset), which is now done in Fourier space.  In the earlier stages, there is an added option to fit a polynomial to the time slices to obtain classical non-linearity coefficients.}
    \label{fig:flowchart}
\end{figure}

The {\sc Solid-waffle} code in ``Advanced'' characterization mode is attempting to measure two types of parameters simultaneously: the conventional parameters (charge per time step $I\Delta t$, gain $g$, classical non-linearity $\beta$, and 3 IPC parameters $\alpha_{\rm H}$, $\alpha_{\rm V}$, and $\alpha_{\rm D}$), which are determined in {\tt pyirc.polychar}; and the $5\times 5$ IPNL kernel parameters $[K^2a+KK^I]_{\Delta x_1,\Delta x_2}$ for $\Delta x_1,\Delta x_2=-2..+2$, which are determined in {\tt pyirc.bfe} (see Fig.~\ref{fig:flowchart}), for a total of 31 parameters. {\sc Solid-waffle} alternates between the two functions, with the IPNL parameters fixed when the conventional parameters are determined in {\tt pyirc.polychar} and the conventional parameters fixed when the IPNL parameters are determined in {\tt pyirc.bfe}.

An updated formula for $C_{abcd}(\Delta x_1,\Delta x_2)$ is relatively straightforward to incorporate. A different strategy is used for the conventional parameters and for the IPNL kernel. The methodology described here is appropriate for the case where IPNL is dominated by BFE, since we established in Paper II that BFE rather than NL-IPC is the dominant form of IPNL in the WFIRST detectors. (We have re-made Figure 6 of Paper II, which examines the slopes of the raw gains and nearest-neighbor correlation functions as a function of signal level, for the three flight candidate SCAs. All three of them show the preference for the ``pure BFE'' instead of ``pure NL-IPC'' model; see \S\ref{sss:corr-results}.) This means that we set $K^I=0$, and can compute the BFE kernel $a$ from $K^2a$ by division in Fourier space.

Computation of $C_{abcd}(\Delta x_1,\Delta x_2)$ also requires the full set of non-linearity coefficients. {\sc Solid-waffle} was originally designed to include and fit $\beta_2$ (previously $\beta$), but since at each iterative step in the parameter determination the gain is known, the higher-order coefficients can be determined from Eq.~(\ref{eq:bcg}).

To implement the Fourier-space calculations described in \S\ref{sec:extension}, we include {\tt ftsolve.solve\_corr}, a routine that computes equation \ref{eqn:fullcorr} using the fast Fourier transform methods of the {\tt numpy} package. For more details, see Fig. \ref{fig:flowchart}.

\subsubsection{Incorporation into conventional parameter determination}

The conventional parameter determination in {\tt pyirc.polychar} is based on Eq.~(I/80). It uses time slices over the range $a..d$, and depends on two integers $\mu'<\mu$ (default: $\mu'=1$, $\mu=3$). It solves for the 6 unknowns $\{I\Delta t, g, \beta, \alpha_{\rm H}, \alpha_{\rm V}, \alpha_{\rm D}\}$ using the 6 measurements $\{\Delta\bar V, \bar C_{\rm H}, \bar C_{\rm V}, \bar C_{\rm D}, c_1, c_0\}$. Here, we recall that $\Delta V$ is the difference of variances
\begin{equation}
\Delta\bar V = \bar C_{a,a+\mu,a,a+\mu,[d-a-\mu]}(0,0)
-\bar C_{a,a+\mu',a,a+\mu',[d-a-\mu]}(0,0);
\label{eq:DVC}
\end{equation}
as usual for variances, for an ideal detector this is $\propto I\Delta t/g^2$. The inter-pixel correlations are
\begin{equation}
\Delta\bar C_{\rm H} = \bar C_{a,a+\mu,a,a+\mu,[d-a-\mu]}(\pm 1,0)
\end{equation}
for the horizontal direction, and similarly for the vertical ($\Delta \bar C_{\rm V}$) and diagonal ($\Delta\bar C_{\rm D}$) directions. The parameters $c_0$ and $c_1$ are linear fits to the median differences $M_{j,j+1}$ between frames $j$ and $j+1$: $M_{j,j+1} = c_0 + c_1 j + \,$residuals.

In the previous version of {\tt pyirc.polychar}, the 4 correlation-based measurements were written in the form of, e.g.,
\begin{equation}
\Delta \bar V = \Delta \bar V_{\rm base} + {\rm Err}[\Delta\bar V],
\end{equation}
where $\Delta \bar V_{\rm base}$ is the formula for $\Delta\bar V$ without IPNL and without higher-order terms in $\beta$ (see Eq.~I/80). We simply replace the formula for ${\rm Err}[\Delta\bar V]$ (see Eq.~I/81) with $\Delta\bar V_{\S\ref{sec:extension}} - \Delta\bar V_{\rm base}$, where $\Delta\bar V_{\S\ref{sec:extension}}$ is the formula based on Eq.~(\ref{eq:DVC}) and the correlation function formula $\bar C_{abcd[n]}(\Delta x_1,\Delta x_2)$ in \S\ref{sec:extension}. We do similar replacements for $\bar C_{\rm H}$, $\bar C_{\rm V}$, and $\bar C_{\rm D}$.

\subsubsection{Incorporation into IPNL determination}

The previous version of determining the IPNL kernel uses the non-overlapping correlation function at times $a<b<c<d$ -- see Eq.~(I/58). In the old system, we computed the kernel as
\begin{equation}
[K^2a+KK^I]_{\Delta x_1,\Delta x_2} = \frac{g^2}{I^2t_{ab}t_{cd}} C_{abcd}(-\Delta x_1,-\Delta x_2) + \left\{ \begin{array}{lll}
2(1-8\alpha)\beta & & (\Delta x_1,\Delta x_2) = (0,0), \\
4\alpha_{\rm H}\beta & & (\Delta x_1,\Delta x_2) = (\pm 1,0), \\
4\alpha_{\rm V}\beta & & (\Delta x_1,\Delta x_2) = (0,\pm 1), \\
0 & & {\rm otherwise}. \\
\end{array} \right.
\label{eq:kc1}
\end{equation}
In the new version, we have a function ({\tt ftsolve.many\_corr}) that solves for $C_{abcd}(\Delta x_1,\Delta x_2)$ in terms of $[K^2a+KK^I]_{\Delta x_1,\Delta x_2}$. If the non-overlapping correlation function is computed in a $5\times 5$ region, and we have a $5\times 5$ IPNL kernel, then we have 25 constraints and 25 unknowns. While in this version Eq.~(\ref{eq:kc1}) is no longer accurate, it is close enough to help us write a nonlinear system solver that converges. At each iterative step of the solver, we compute the predicted correlation function\footnote{In practice, we perform most of this computation in Fourier space, as per \S\ref{sec:extension}. To initiate the calculation, we must solve for $a$. As notated in configuration space, $[K^2a+KK^I]$ indicates a convolution of the BFE kernel $a$ with the IPC kernels $K$ and $K^I$, so its Fourier transform may be expressed as products of $\widetilde{K}$, $\widetilde{K}^I$, and $\widetilde{a}$:
\begin{equation}
    \widetilde{a} = \frac{\widetilde{\mathrm{IPNL}}-\widetilde{K}\widetilde{K}^I}{\widetilde{K}^2}.
\end{equation} Then $a$ is obtained by computing the inverse Fourier transform.
} and update the IPNL kernel:
\begin{eqnarray}
[K^2a+KK^I]_{\Delta x_1,\Delta x_2} &+=& 
\frac{g^2}{I^2t_{ab}t_{cd}} C_{abcd}(-\Delta x_1,-\Delta x_2;{\rm measured})
\nonumber \\ && -\frac{g^2}{I^2t_{ab}t_{cd}} C_{abcd}(-\Delta x_1,-\Delta x_2;{\rm predicted}).
\end{eqnarray}
This would converge in one step if Eq.~(\ref{eq:kc1}) were valid; in practice it is usually found to converge rapidly.

\subsection{Tests on simulated data}
\label{ss:sims}

For Paper I, we created a test bed of simulated flats and darks (``Paper I sims") to assess the performance of {\sc Solid-waffle}.  Section 4.2 of Paper I describes the procedure for generating this simulated data set, and we apply a nearly identical pipeline for this paper. Here, we have created a second test bed of simulated flats and darks (10 each) with an updated treatment of the classical non-linearity ingredient. Previously we input only the quadratic coefficient of the classical non-linearity curve ($\beta$), and now we include higher order polynomial coefficients up to the quartic coefficient: $\beta_j$ as given by equation~\ref{eq:bcg}, for $j=2,3,4$.  The input choices for these $\beta_j$ were motivated by fits obtained on SCA 20829: $\beta_2=1.5725 \times 10^{-6}$, $\beta_3=-1.9307 \times 10^{-11}$, and $\beta_4=1.4099 \times 10^{-16}$ with units of electrons$^{1-j}$.\footnote{These are not the same polynomial coefficients that appear in \S\ref{sec:characterization}, because the coefficients chosen for the simulation were based on a fit that goes almost to full well.}  {\sc Solid-waffle} reports these coefficients multiplied by factors of $g^{1-j}$, such that $g\beta_2=3.329 \times 10^{-6}$ DN$^{-1}$, $g^{2}\beta_2=-8.193 \times 10^{-11}$ DN$^{-2}$, and $g^{3}\beta_3=1.233 \times 10^{-15}$ DN$^{-3}$, and thus later results will be reported in the DN-based units.  All other inputs remain the same as for Paper I, so we refer the reader to Paper I/Section 4.2 for further details.  We refer to this simulation data set as ``Paper III sims.''

We ran the latest version of {\sc Solid-waffle}, using fits only to the quadratic coefficient $\beta$, on the Paper I sims to directly compare its performance against the Paper I version of {\sc Solid-waffle}, which dropped higher order terms.  As was found in Paper I, we found the recovered charge per time slice, gain, IPC $\alpha$, and $\beta$ matched the input `truth' values well, with small offsets of less than a percent.  More importantly, the central value at zero-lag is $[K^2a']_{0,0}=-1.1715 \pm 0.0073$ (stat) ppm/e, and the averaged nearest neighbor $[K^2a']_{\langle 1,0\rangle }=0.2004 \pm 0.0036$ (stat) ppm/e for $32\times32$ super-pixels. Compared to the input values of $[K^2a']_{0,0,{\rm input}}=-1.1590$ ppm/e and $[K^2a']_{\langle 1,0\rangle ,{\rm input}}=0.2034$ ppm/e, the biases are now 1.1\% and 1.5\%, compared to the original 12.1\% and 2.7\% biases found in Paper I.  This finding supports the Paper I hypothesis that the likely source of the 12.1\% bias was the exclusion of higher order terms in that analysis.

We also ran the latest version of {\sc Solid-waffle}, with fits to $\beta_j$ up to the quartic coefficient, on the Paper III sims.  Fig.~\ref{fig:paramplot} visualizes the results in the bottom row of panels.  We report on the differences between the inputs and outputs for two different bin choices of super-pixels, $16\times16$ and $32\times32$.  As above, the charge per time slice, gain, IPC alpha, and $\beta_j$ outputs match the input `truth' well.  The central value at zero-lag is $[K^2a']_{0,0}=-1.1574 \pm 0.0073$ (stat) ppm/e  for $16\times16$ super-pixels and $[K^2a']_{0,0}=-1.1701 \pm 0.0148$ (stat) ppm/e for $32\times32$ super-pixels.  Compared to the input value $[K^2a']_{0,0,{\rm input}}=-1.1590$ ppm/e, this translates to biases of 0.0016 ppm/e (0.1\%, 0.22$\sigma$) and 0.0111 ppm/e (0.96\%, 0.75$\sigma$), respectively.  These biases are again much less than the 12.1\% bias in the recovered central BFE coefficient in Paper I. We note that using more super-pixels (i.e. the $32\times32$ binning) gives slightly larger biases in the recovered parameters.  Examining $[K^2a']_{0,0}$, for example, we see that the distribution of values over super-pixels is skewed, which suggests a noise rectification bias.  Since the correlation function is a non-linear function of the input parameters, noise fluctuations that pull the fit parameters up do not cancel those that pull them down.

For the symmetrically averaged nearest neighbors, we found $[K^2a']_{\langle 1,0\rangle }=0.2010 \pm 0.0036$ (stat) ppm/e for $16\times16$ superpixels and $[K^2a']_{\langle 1,0\rangle }=0.2003 \pm 0.0073$ (stat) ppm/e for $32\times32$ superpixels.  Compared to the input value $[K^2a']_{\langle 1,0\rangle ,{\rm input}}=0.2034$ ppm/e, this translates to biases of 0.0024 ppm/e (1.2\%, 0.67$\sigma$) and 0.0031 ppm/e (1.5\%, 0.42$\sigma$).  These are again smaller than the 2.7\% bias found for the nearest neighbors in Paper I.

\section{Application to detector characterization data}
\label{sec:characterization}

We now apply our improved formalism to laboratory data from {\slshape WFIRST} development and flight candidate detectors, and comment on the consistency of results and the features that we identified during analysis.

\subsection{Description of the data}

Data for this paper were acquired at the Detector Characterization Laboratory (DCL) at the NASA Goddard Spaceflight Center. We have used some of the older flat/dark data from the development detector SCA 18237 that was used in Paper II, for ease of comparison with previous results. However, we have now turned most of our attention to the flight candidate detectors; this paper analyzes three of these, SCAs 20663, 20828, and 20829.

\begin{figure}
    \centering
    \includegraphics[width=3.5in]{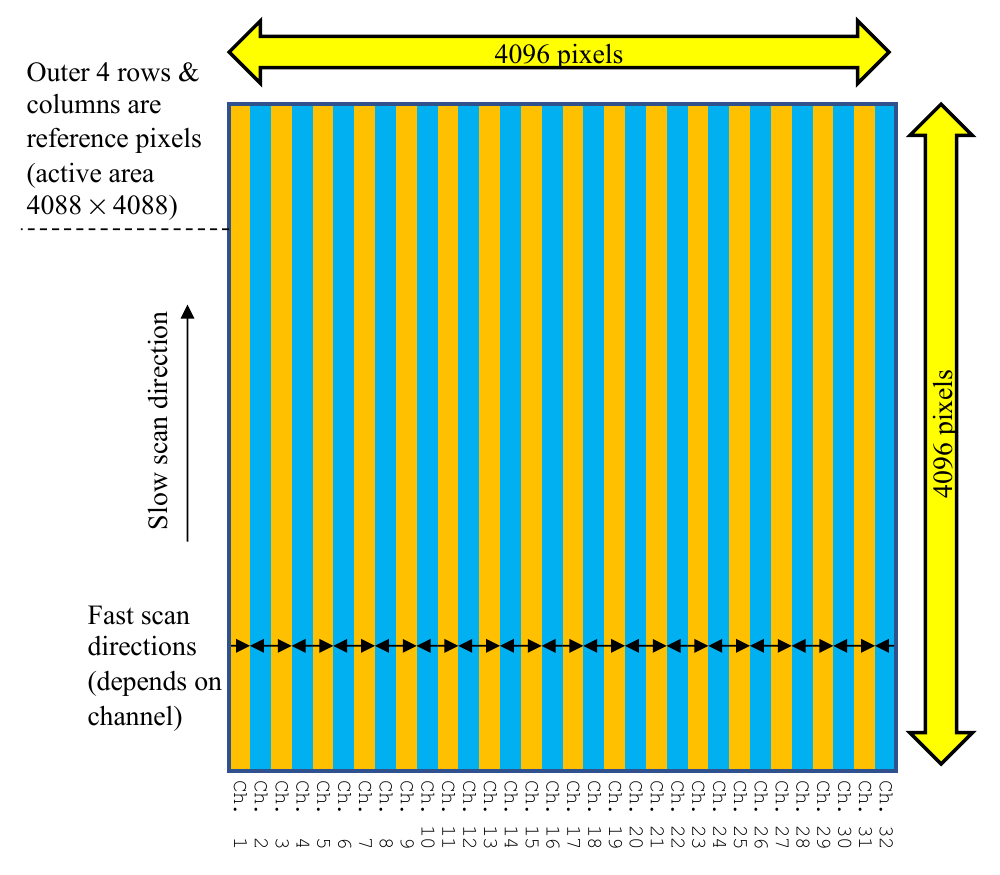}
    \caption{Readout of an H4RG with 32 output channels, with the image in the usual display orientation for FITS files. Each channel is 128 columns wide and 4096 rows high. Pixels are read within each row in the horizontal (fast scan) direction, with even and odd outputs being read in opposite directions. The rows are read in sequence, starting from the bottom (row 0) and going to the top (row 4095). The detector overall is $4096\times 4096$ pixels, but because the outer 4 rows and columns are reference pixels, the active area is only $4088\times 4088$.}
    \label{fig:h4rgfig}
\end{figure}

\begin{figure}
    \centering
    \includegraphics[width=6.5in]{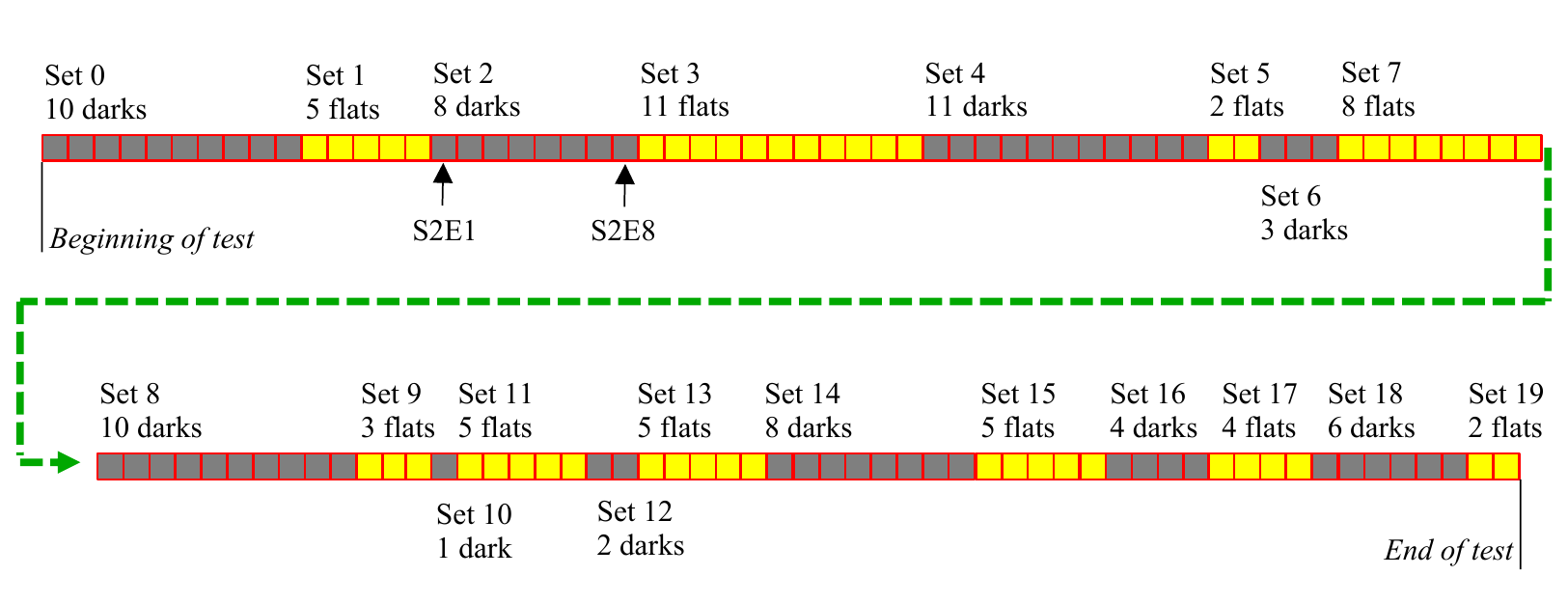}
    \caption{The sequence of flats and darks at 1.4 $\mu$m used for SCAs 20663, 20828, and 20829.}
    \label{fig:se}
\end{figure}

Data for SCAs 20663, 20828, and 20829 were acquired during acceptance testing. All of these data were acquired with a Leach controller, with 32 output channels (each 128 columns wide), at 1.0 V bias, and with the guide window off.\footnote{Acceptance testing includes tests of the guide window, and at other bias voltages, but that data are not used in this paper.} The readout pattern and 32 channels are shown in Fig.~\ref{fig:h4rgfig}. We have mostly used the flat/dark sequences. This sequence consists of ``sets'' of exposures; the even-numbered sets are darks and the odd-numbered sets are flats. Each set contains a variable number of exposures, as shown in Figure~\ref{fig:se}; we denote, e.g., the 4th exposure in the 2nd set is S2E4. Note that set numbers start at 0 (i.e., with darks) but exposure numbers start at 1. Each exposure contains 64 non-destructively read frames (not counting the initial reset frame), with each frame having a time duration of 2.764 s for a total duration of $64\times 2.764 = 176.896$ s. The flats considered here are taken at a wavelength of 1.4 $\mu$m in order to avoid quantum yield effects. The flux is typically $\sim 1200$ e/p/s, resulting in the SCAs reaching full well about half way through a flat field exposure and sitting in saturation for $\sim 90$ s before they reset. This results in significant persistence over the course of the test. It also results in a change in response of the detector to light following a previous exposure (``burn-in'') that is not simply the additive effect of the persistence.
The persistence and burn-in effects show similar spatial structures and presumably have a related physical origin due to charge traps \citep[e.g.,][]{2008SPIE.7021E..0JS, 2012SPIE.8442E..4WR, 2018SPIE10709E..1BR}; however we have not attempted to model or correct these effects in our analysis.

We have also used some lower-intensity flats for each SCA ($\sim 300$ e/p/s; 11 frames, 10 exposures) taken for low-signal gain determination, paired with some darks taken for noise measurements. This enables us to test whether the gain and IPC extrapolated from higher signal levels apply to the low signal levels; however these low signal levels are not as good for probing the BFE.

Finally, the DCL has provided single pixel reset (SPR) data as a $2\times 4096\times 4096$ cube. This data contains one frame taken following a reset, and then a second frame after every 8th pixel in $x$ and 8th pixel in $y$ (i.e., approximately 1/64 of the pixels in total\footnote{When we take into account the reference pixels and boundary effects between readout channels, in fact 245280 instead of $(4096/8)^2=262144$ pixels are reset.}) has been reset to a higher voltage. Since the charge changed only in the reset pixel, this method can be used to determine the IPC kernel; the spatial resolution is better than the flat field autocorrelation method and it can distinguish, e.g., up vs.\ down ($K_{0,1}$ vs.\ $K_{0,-1}$).

The flat/dark sequence data for the development detector SCA 18237 was described in Paper II, but the major differences include: (i) 64-channel readout (which we will not use in flight); (ii) the lower lamp intensity, such that saturation occurs near the end of the exposure; (iii) the wavelength of light used was $1.2$ $\mu$m instead of 1.4 $\mu$m; (iv) the data were acquired in a different dewar; (v) 66 instead of 64 frames; and (vi) the sequence of flats and darks is somewhat different (and is numbered differently: in the development tests, the initial set of darks is Set 1 instead of Set 0). 

\subsection{Application to a development detector (SCA 18237)}

\begin{table}[]
\scriptsize  
    \centering
    \begin{tabular}{|ll|*{6}{c}|}
    \hline\hline
    \multicolumn{8}{|c|}{SCA 18237, $\beta_{2}$ fit} \\
    \hline
    \multirow{2}{*}{} & &\multicolumn{3}{c}{1st,n3} &\multicolumn{3}{c|}{fid,n23}\\
    Quantity     &Units &short &fid &long &short &fid &long \\
    \hline
     
    $[K^{2}a + KK^I]_{0,0}$   &ppm/e     &-1.3681 &-1.4713 &-1.6668 &- &- &- \\
    $\sigma([K^{2}a + KK^I]_{0,0})$   &ppm/e     &0.0334 &0.0181 &0.0108 &- &- &- \\
    $[K^{2}a + KK^I]_{\langle 1,0\rangle }$ &ppm/e     &0.2373 &0.2233 &0.2432 &- &- &- \\
    
    $\sigma([K^{2}a + KK^I]_{\langle 1,0\rangle })$ &ppm/e     &0.0165 &0.0083 &0.0043 &- &- &- \\
    \hline
    \multicolumn{8}{|c|}{SCA 18237, $\beta_{i}$ fit, i=2,3,4} \\
    \hline
        $[K^{2}a + KK^I]_{0,0}$   &ppm/e     &-1.3610 &-1.3982 &-1.5674 &-1.0858 &-1.2486 &-1.4868  \\
    $\sigma([K^{2}a + KK^I]_{0,0})$   &ppm/e     &0.0333 &0.0181 &0.0101 &0.0092 &0.0052 &0.0060 \\
    $[K^{2}a + KK^I]_{\langle 1,0\rangle }$ &ppm/e     &0.2371 &0.2252 &0.2447 &0.2300 &0.2314 &0.2469  \\
    
    $\sigma([K^{2}a + KK^I]_{\langle 1,0\rangle })$ &ppm/e     &0.0165 &0.0083 &0.0043 &0.0045 &0.0024 &0.0012 \\
    \hline
    \end{tabular}
    \caption{Central and nearest neighbor BFE coefficients for the SCA 18237 data from Paper II using the updated formalism from this work.  ``1st,n3" and ``fid,n23" refer to sets of flats used in Paper II, and the number refers to the number of flats used in the respective cases. ``short", ``fid", and ``long" refer to the baseline frame choices where $abcd=$ (3 7 9 13), (3 11 13 21), and (3 19 21 37) respectively. `-' is given for values we did not compute.}
    \label{tab:sca18237_comp}
\end{table}
    
Here, we compare measurements for SCA 18237 from Paper II with results using the updated {\sc Solid-waffle}.  We focus on measurements for the ``1st,n3" (3 flats taken immediately after darks) and ``fid,n23" cases (23 flats taken at various times after darks), and for the cases of fitting only the quadratic $\beta_{2}$ (known simply as $\beta$ in Paper II) and fitting up to the quartic $\beta_j$.  The frame choice of 3,11,13,21 is the same as the fiducial set used in Paper II.  First, for the ``1st,n3" flats, both code runs return nearly identical values of charge per time step, gain, IPC $\alpha$ as those presented in Paper II.  When the quadratic term $\beta$ is fit, we obtain $\beta=0.5829 \pm 0.0010$ ppm/e, comparable to the Paper II value of $\beta=0.5830 \pm 0.0010$. We summarize the central and averaged nearest neighbor coefficients in Table~\ref{tab:sca18237_comp}.  In Table~\ref{tab:sca18237_comp}, this frame choice of $abcd=3,11,13,21$ is denoted as `fid'. The central BFE coefficient $[K^2a+KK^I]_{0,0}=-1.4713 \pm 0.0181$ ppm/e is significantly less than the $[K^2a+KK^I]_{0,0}=-1.2004 \pm 0.0154$ ppm/e of Paper II, with the Paper II value being 18.4\% higher. The averaged nearest neighbor is higher than that of Paper II by 4\%. When the $\beta_j$ are fit up to the quartic term, we see similar comparisons such that the Paper I values are 14.1\% higher and 4.9\% lower, respectively.

In Paper II, we ran alternative time intervals and found larger central and nearest neighbor IPNL values for shorter time baselines and smaller values for longer time baselines. We ran the same choices of time intervals with the updated {\sc Solid-waffle} and report on the results for both the ``1st,n3" and ``fid,n23" sets of flats.  Note that for the ``fid,n23" flats, we only ran the code with the configuration that fits $\beta_j$ up to the quartic term, as these runs take significantly more computing time, and we have already compared code configurations with  ``1st,n3". In both Paper II and here, we find the central IPNL coefficient increases when the time baseline increases.   Here, as the values in Table~\ref{tab:sca18237_comp} show, we find that the nearest neighbor IPNL coefficient now trends slightly upward going to longer baselines (by 0.0169 ppm/e for ``fid,n23"), as opposed to the Paper II trend in which the nearest neighbor coefficient decreased going to longer baselines (by 0.0274 ppm/e).

\subsection{Application to flight candidate detectors (SCAs 20663, 20828, and 20829)} \label{ss:application}

We now apply our machinery to the first three flight candidate detector arrays for {\slshape WFIRST}: SCAs 20663, 20828, and 20829. All three SCAs are different, however we will show most results for only one SCA for reasons of space. We choose SCA 20829 since it has the median performance in terms of offset between single pixel reset and autocorrelation IPC measurements (SCA 20663 shows larger offsets and SCA 20828 shows smaller offsets).

\subsubsection{General properties of the data}

\begin{figure}
    \centering
    \includegraphics[width=2.1in]{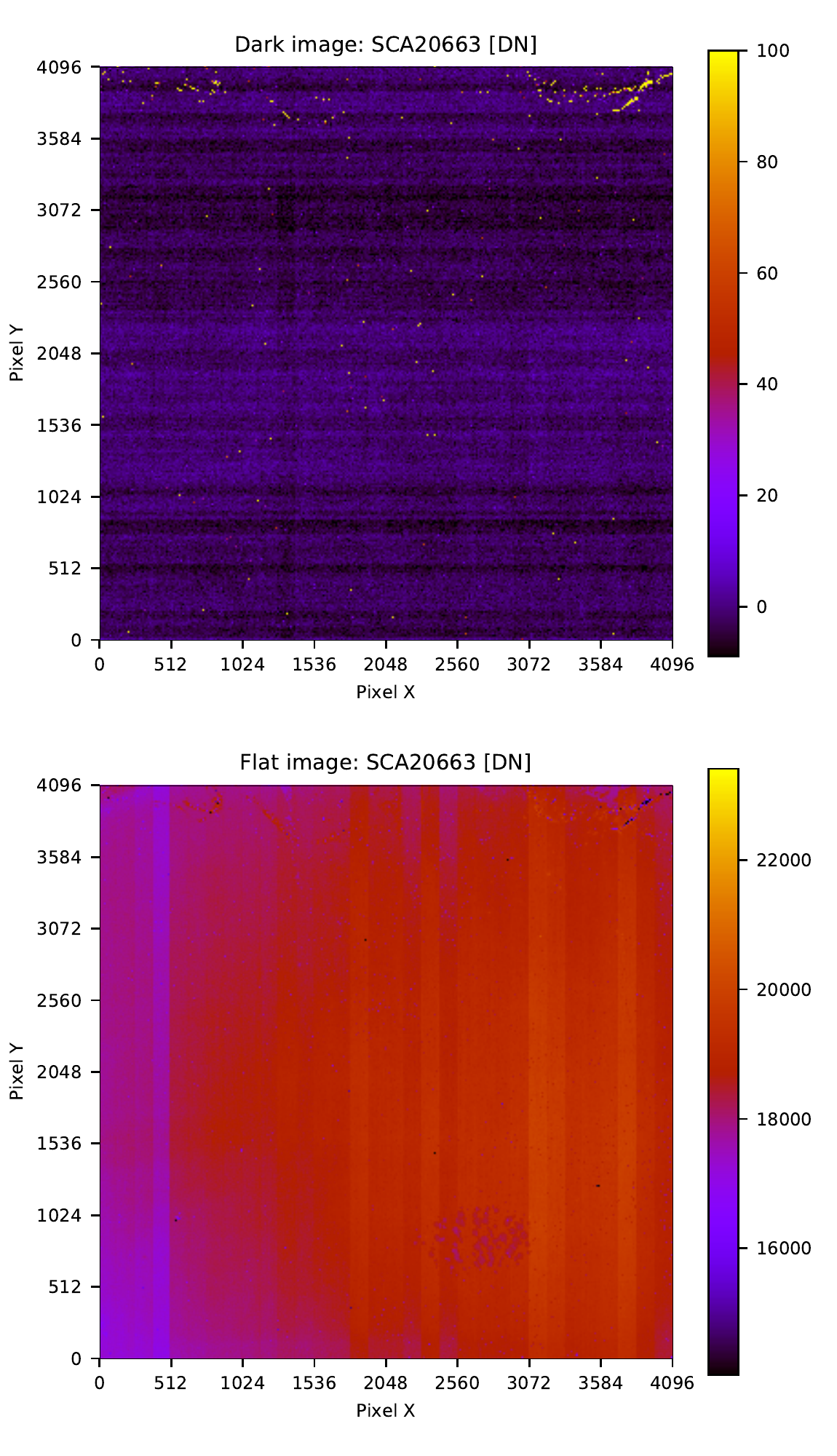}
    \includegraphics[width=2.1in]{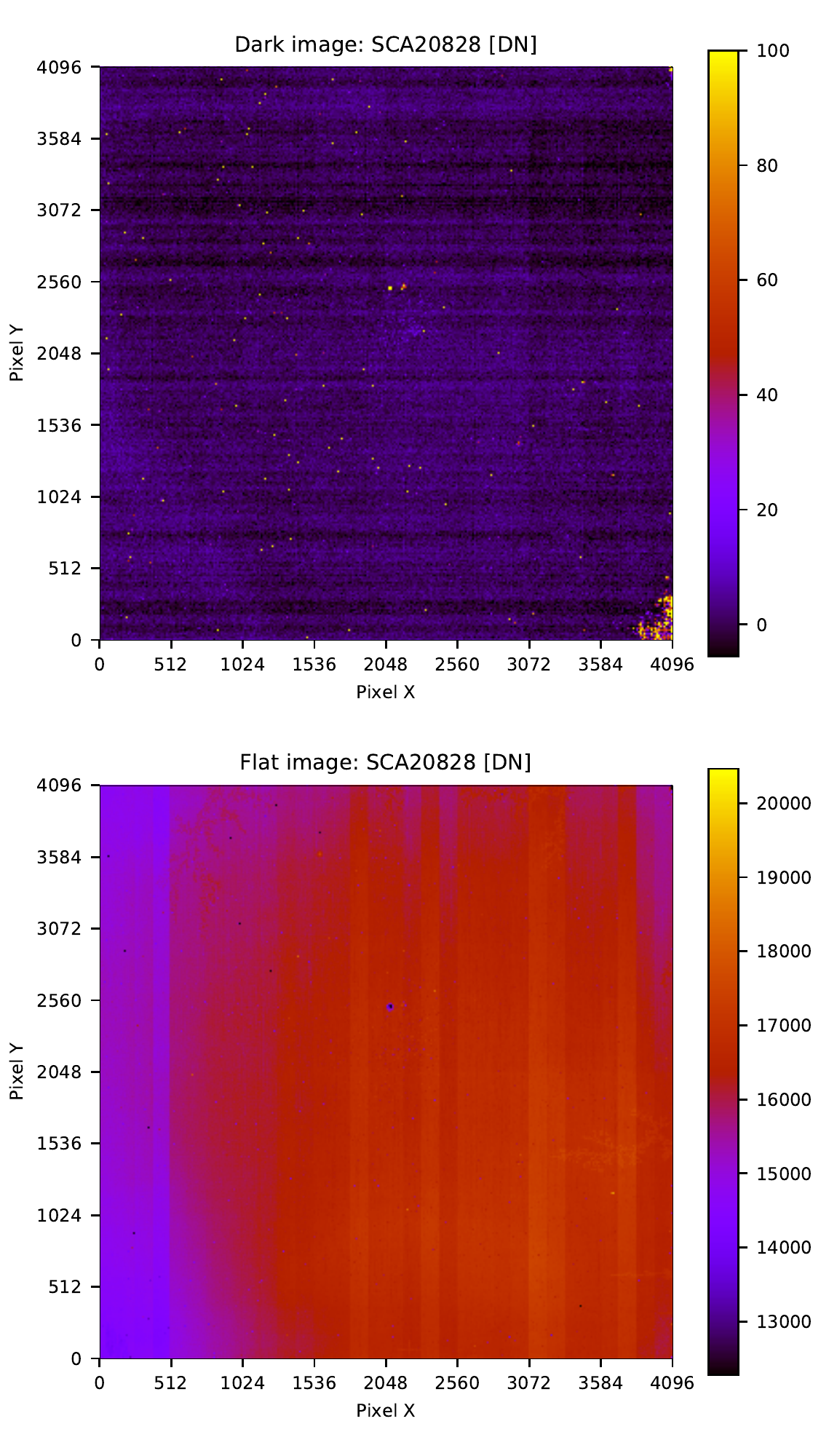}
    \includegraphics[width=2.1in]{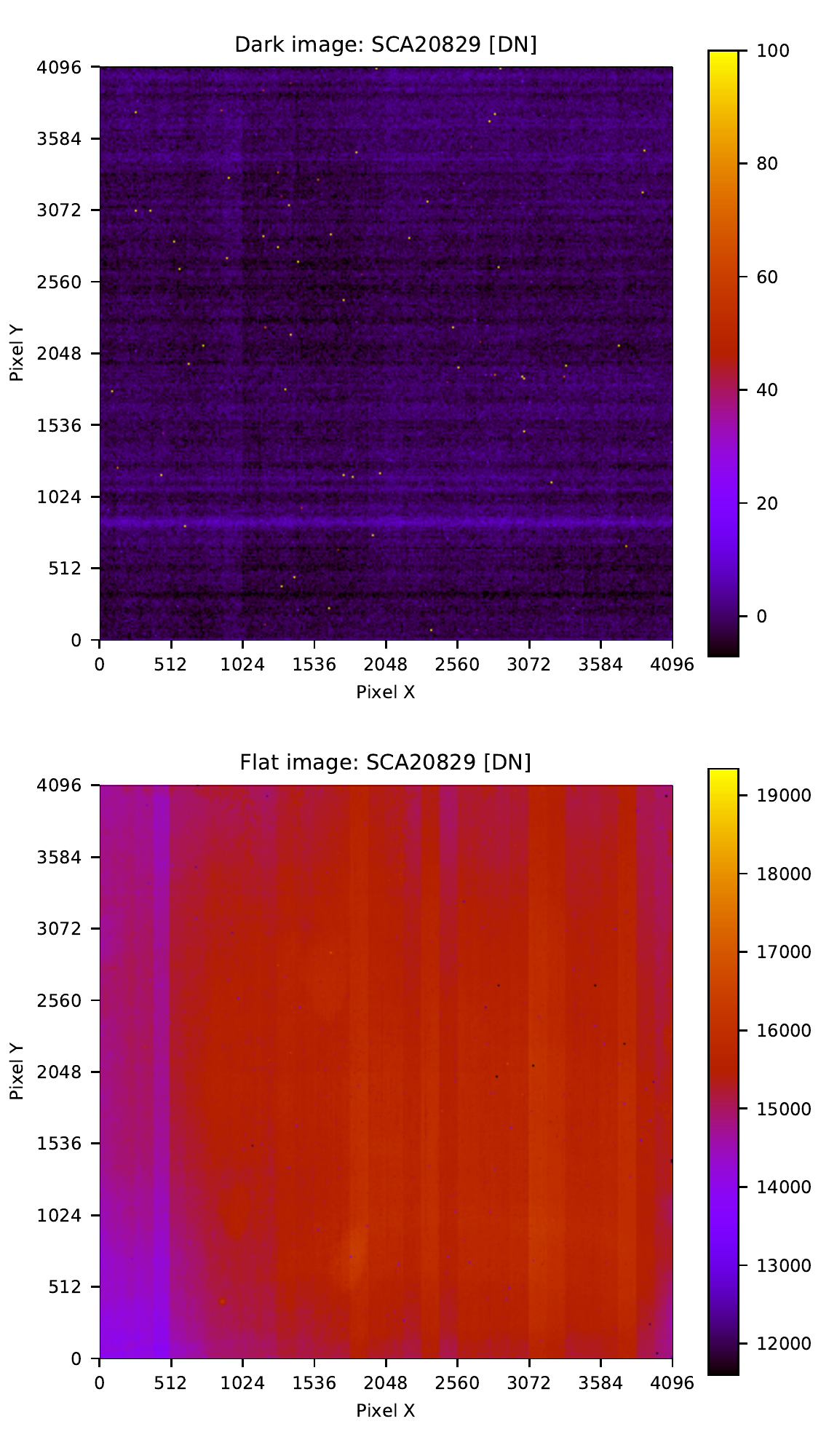}
    \caption{A dark frame (S0E1; top) and a flat frame (S1E1; bottom), for SCAs 20663 (left), 20828 (middle), and 20829 (right). The dark and flat are averaged into $4\times 4$ pixel bins. All data presented here are based on the difference between the 1st and 11th frames, $S_{1,11}$. No reference pixel subtraction has been used. Some isolated cosmetic defects can be seen, especially on the top of SCA20663 and the lower-right corner of SCA20828.}
    \label{fig:dfs}
\end{figure}

Example dark and flat images of SCAs 20663, 20828, and 20829 are shown in Fig.~\ref{fig:dfs}. The cosmetic quality of the detectors is excellent.

\begin{figure}
    \centering
    \includegraphics[width=3.1in]{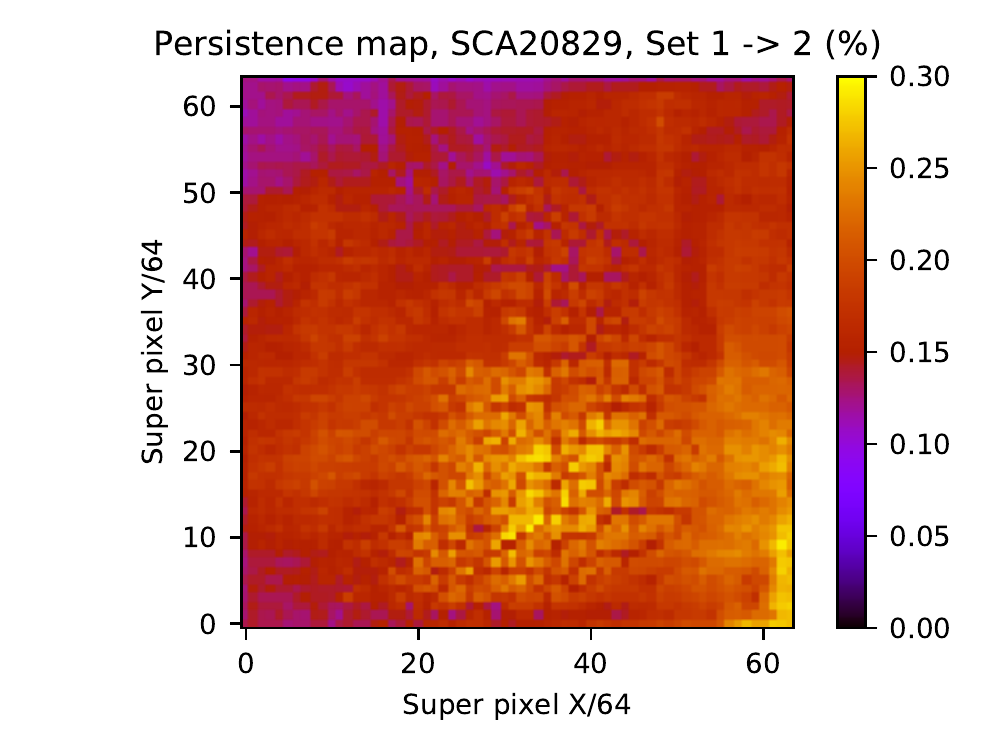}
    \includegraphics[width=3.1in]{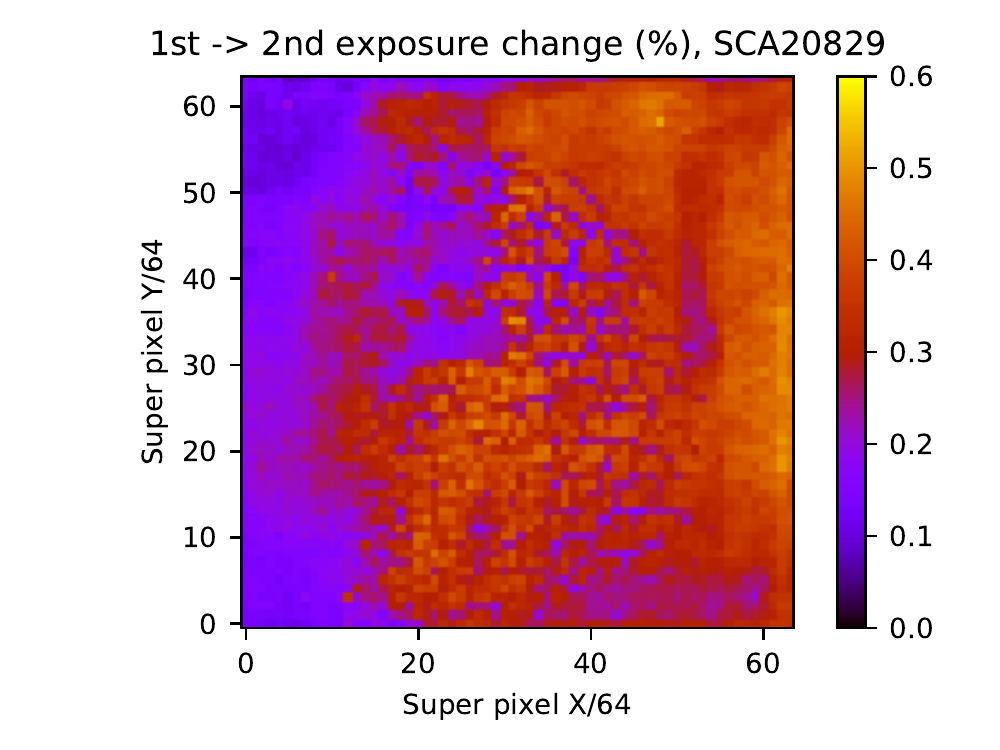}
    \caption{Images of the persistence and burn-in in SCA 20829. Both images are based on the signal map $S_{1,10}$ dark and flat data binned into 4096 $64\times 64$ superpixels. The left panel is the persistence, based on the ratio of S2E1/S1E1. The right panel is the burn-in, measured as the second frame effect, (S$n$E2-S$n$E1)/S1E1, averaged over flat sequences ($n=1,3,5,...19)$. Note that some of the same spatial structures appear in both maps.}
    \label{fig:ph}
\end{figure}

Examination of the dark and flat sequences shows some evidence for previously noted behaviors, including persistence and burn-in, as shown in Fig.~\ref{fig:ph}. The persistence signal (left frame) is measured based on the ratio of $S_{1,10}$ (i.e., the first 25 s of the exposure) in S2E1 (the first dark following a sequence of 5 flats) to that in S1E1. Of the 4096 super-pixels (each $64\times 64$ pixels) shown in the figure, the median persistence is 0.17\% of the initial flat field (the full range is 0.10 to 0.29\%). In the second dark exposure, S2E2, this median persistence has dropped to 0.01\% (full range $-$0.01 to 0.04\%). We also show the burn-in, as measured by the change in response going from the 1st to 2nd exposure in a flat set. The median change in response in the 2nd exposure in the 4096 super-pixels is 0.29\% (the full range is 0.08 to 0.52\%). Both maps show some of the same spatial structure. As a result of the persistence, we have not used first darks in the correlation analysis.

\subsubsection{Correlation-based results}
\label{sss:corr-results}

\begin{figure}[hbtp]
    \centering
    \includegraphics[width=6.5in]{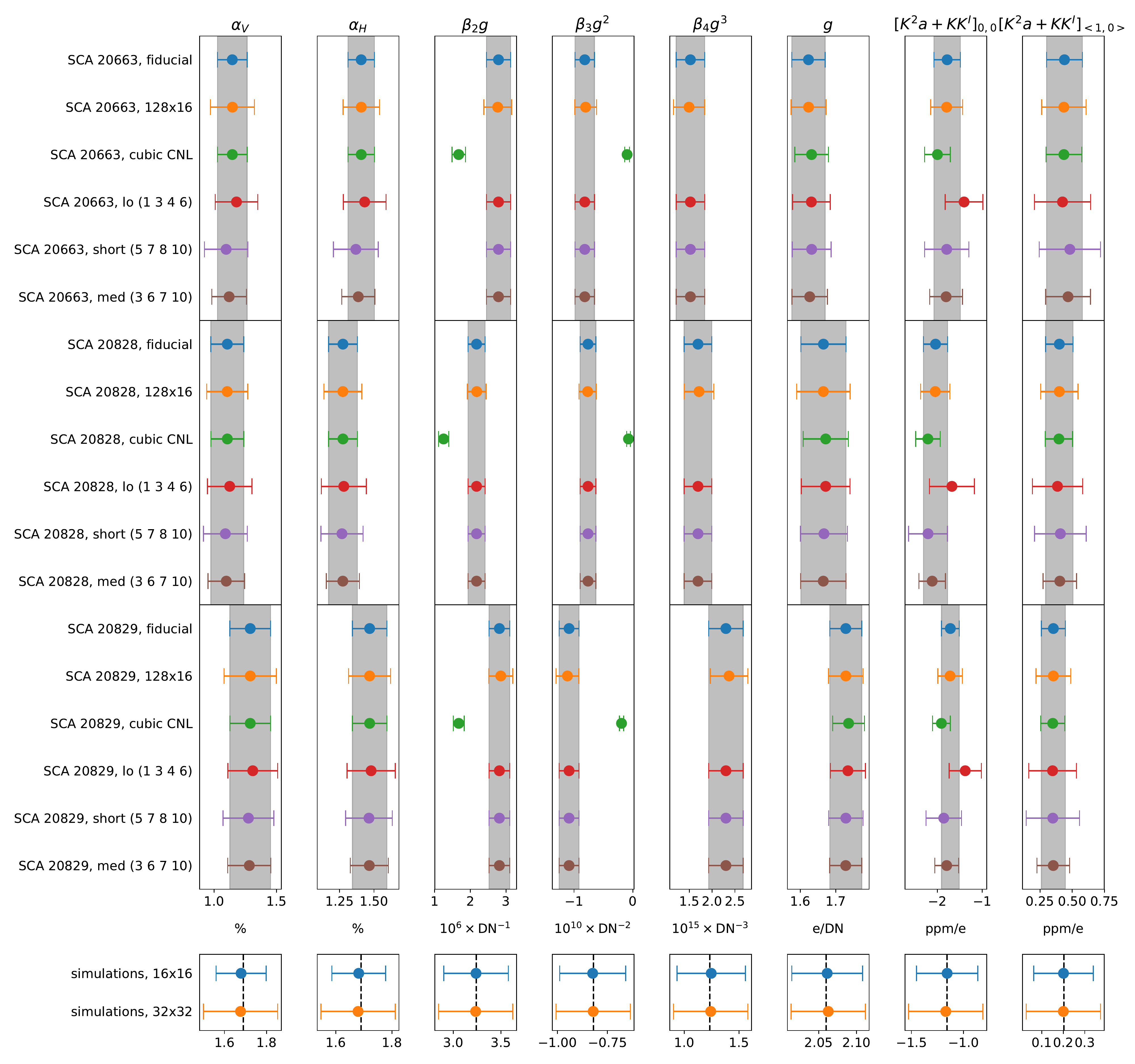}
    \caption{IPNL and CNL parameters for SCA 20663, SCA 20828, SCA 20829, and simulated data, obtained with various configuration settings as noted on the left axis. Error bars indicate one standard deviation, i.e. 32 times the error on the mean (or $32\sqrt{2}$ times the error on the mean, for the 128x16 configuration). Gray bands are an eye guide corresponding to the fiducial configuration for each SCA. Dashed black lines are true values for the simulated flats. The final column, $[K^2a+KK^I]_{\langle 1,0\rangle }$, indicates the average of the four (horizontal and vertical) ``nearest neighbors'' to the central pixel of the IPNL kernel.}
    \label{fig:paramplot}
\end{figure}

The data from each SCA is processed through our machinery with the following fiducial configuration:
\begin{itemize}
    \item Results are averaged over the 16 flats from Set 1 and Set 3 (see Fig. \ref{fig:dfs}), paired with darks from Set 0 and exposures 2--7 of Set 2. Exposure 1 from Set 2 is not used in order to minimize persistence effects.
    \item Superpixels are defined in a $32\times32$ configuration, such that each superpixel is $128\times128$ pixels.
    \item Frames 1, 5, 6, and 10 are chosen for timeslices $abcd$.
    \item Advanced characterization is set to perform 3 iterations (ncycle=3).
    \item The CNL curve is fit to 4th order, per \S\ref{ss:cnl-curve}, from frame 1 to frame 10.
\end{itemize}
We also process the data using variants on the fiducial configuration, including
\begin{itemize}
    \item binning into $128\times16$ superpixels;
    \item fitting the CNL curve to cubic order;
    \item clustering timeslices near the beginning (1 3 4 6) or end (5 7 8 10) of the fiducial timeslice window, and eliminating the earliest part of the fiducial window (3 6 7 10); and
    \item changing the $\epsilon$ parameter from Paper I (percentile used to reject outliers in the correlation function analysis) to 0.02 instead of the default 0.01.
\end{itemize}
The configuration for processing simulated data is described in \S\ref{ss:sims}.

\begin{table}[]
\scriptsize  
    \centering
    \begin{tabular}{ll|ccccccc}
    \hline\hline
    Quantity     &Units &fid &128x16 &cubic &lo &short &med &$\epsilon=0.02$ \\
    \hline
    \multicolumn{9}{c}{SCA 20663} \\
    \hline
    Charge, $It_{n,n+1}$     &ke                        &3.5055 &3.5054 &3.5027 &3.5234 &3.5240 &3.5118 &3.5045 \\
    Gain $g$                 &e/DN                      &1.6232 &1.6234 &1.6319 &1.6314 &1.6322 &1.6267 &1.6227 \\
    IPC $\alpha$             &\%                        &1.2762 &1.2771 &1.2763 &1.3058 &1.2323 &1.2531 &1.2761 \\
    IPC $\alpha_{\rm H}$     &\%                        &1.4070 &1.4075 &1.4070 &1.4323 &1.3683 &1.3859 &1.4071 \\
    IPC $\alpha_{\rm V}$     &\%                        &1.1455 &1.1467 &1.1456 &1.1793 &1.0963 &1.1202 &1.1451 \\
    IPC $\alpha_{\rm D}$     &\%                        &0.1210 &0.1219 &0.1212 &0.1172 &0.0953 &0.1086 &0.1216 \\
    $\beta_2g$               &$10^6\times$DN$^{-1}$     &2.7899 &2.7674 &1.6752 &2.7899 &2.7899 &2.7899 &2.7899 \\
    $\beta_3g^2$             &$10^{10}\times$DN$^{-2}$  &-0.8173 &-0.8027 &-0.0978 &-0.8173 &-0.8173 &-0.8173 &-0.8173 \\
    $\beta_4g^3$             &$10^{15}\times$DN$^{-3}$  &1.5245 &1.4948 &N/A &1.5245 &1.5245 &1.5245 &1.5245 \\

    $[K^{2}a + KK^I]_{0,0}$   &ppm/e     &-1.7800 &-1.7914 &-1.9967 &-1.4029 &-1.7899 &-1.8033 &-1.7957 \\
    $[K^{2}a + KK^I]_{\langle 1,0\rangle }$ &ppm/e     &0.4396 &0.4348 &0.4353 &0.4245 &0.4816 &0.4673 &0.4273 \\
    $[K^{2}a + KK^I]_{\langle 1,1\rangle }$ &ppm/e     &0.1559 &0.1526 &0.1558 &0.1611 &0.2008 &0.1776 &0.1442 \\
    $[K^{2}a + KK^I]_{\langle 2,0\rangle }$ &ppm/e     &0.0707 &0.0665 &0.0709 &0.0870 &0.1020 &0.0884 &0.0610 \\
    $[K^{2}a + KK^I]_{\langle 2,1\rangle }$ &ppm/e     &0.0477 &0.0443 &0.0479 &0.0561 &0.0700 &0.0596 &0.0414 \\
    $[K^{2}a + KK^I]_{\langle 2,2\rangle }$ &ppm/e     &0.0338 &0.0302 &0.0339 &0.0473 &0.0621 &0.0469 &0.0267 \\
    $[K^{2}a+KK^I]_{\rm H}$   &ppm/e     &0.4329 &0.4289 &0.4280 &0.4184 &0.4748 &0.4601 &0.4197 \\
    $[K^{2}a+KK^I]_{\rm V}$   &ppm/e     &0.4462 &0.4408 &0.4425 &0.4306 &0.4884 &0.4746 &0.4349 \\
    
    \hline
    \multicolumn{9}{c}{SCA 20828} \\
    \hline
    Charge, $It_{n,n+1}$     &ke                        &3.0869 &3.0864 &3.0852 &3.0987 &3.0894 &3.0856 &3.0878 \\
    Gain $g$                 &e/DN                      &1.6654 &1.6651 &1.6717 &1.6716 &1.6669 &1.6649 &1.6659 \\
    IPC $\alpha$             &\%                        &1.1900 &1.1892 &1.1900 &1.2027 &1.1787 &1.1854 &1.1874 \\
    IPC $\alpha_{\rm H}$     &\%                        &1.2743 &1.2737 &1.2744 &1.2806 &1.2673 &1.2734 &1.2716 \\
    IPC $\alpha_{\rm V}$     &\%                        &1.1056 &1.1047 &1.1057 &1.1249 &1.0901 &1.0974 &1.1031 \\
    IPC $\alpha_{\rm D}$     &\%                        &0.1317 &0.1306 &0.1318 &0.1283 &0.1277 &0.1349 &0.1310 \\
    $\beta_2g$               &$10^6\times$DN$^{-1}$     &2.1758 &2.1799 &1.2520 &2.1758 &2.1758 &2.1758 &2.1758 \\
    $\beta_3g^2$             &$10^{10}\times$DN$^{-2}$  &-0.7621 &-0.7688 &-0.0740 &-0.7621 &-0.7621 &-0.7621 &-0.7621 \\
    $\beta_4g^3$             &$10^{15}\times$DN$^{-3}$  &1.6909 &1.7134 &N/A &1.6909 &1.6909 &1.6909 &1.6909 \\

    $[K^{2}a + KK^I]_{0,0}$   &ppm/e     &-2.0386 &-2.0451 &-2.2090 &-1.6736 &-2.2057 &-2.1132 &-2.0371 \\
    $[K^{2}a + KK^I]_{\langle 1,0\rangle }$ &ppm/e     &0.3999 &0.3997 &0.3966 &0.3856 &0.4082 &0.4038 &0.3990 \\
    $[K^{2}a + KK^I]_{\langle 1,1\rangle }$ &ppm/e     &0.1069 &0.1077 &0.1066 &0.1115 &0.1162 &0.0996 &0.1071 \\
    $[K^{2}a + KK^I]_{\langle 2,0\rangle }$ &ppm/e     &0.0271 &0.0267 &0.0271 &0.0416 &0.0211 &0.0177 &0.0260 \\
    $[K^{2}a + KK^I]_{\langle 2,1\rangle }$ &ppm/e     &0.0076 &0.0080 &0.0076 &0.0109 &0.0017 &0.0033 &0.0073 \\
    $[K^{2}a + KK^I]_{\langle 2,2\rangle }$ &ppm/e     &0.0019 &0.0010 &0.0019 &0.0136 &-0.0006 &-0.0054 &0.0028 \\
    $[K^{2}a+KK^I]_{\rm H}$   &ppm/e     &0.3814 &0.3800 &0.3777 &0.3768 &0.3814 &0.3797 &0.3800 \\
    $[K^{2}a+KK^I]_{\rm V}$   &ppm/e     &0.4185 &0.4194 &0.4155 &0.3943 &0.4350 &0.4278 &0.4179 \\
    \hline
    
    \multicolumn{9}{c}{SCA 20829} \\
    \hline
    Charge, $It_{n,n+1}$     &ke                        &3.0515 &3.0520 &3.0496 &3.0623 &3.0513 &3.0504 &3.0524 \\
    Gain $g$                 &e/DN                      &1.7285 &1.7285 &1.7362 &1.7344 &1.7286 &1.7282 &1.7290 \\
    IPC $\alpha$             &\%                        &1.3790 &1.3789 &1.3789 &1.3945 &1.3692 &1.3741 &1.3773 \\
    IPC $\alpha_{\rm H}$     &\%                        &1.4684 &1.4681 &1.4683 &1.4791 &1.4635 &1.4659 &1.4662 \\
    IPC $\alpha_{\rm V}$     &\%                        &1.2896 &1.2897 &1.2896 &1.3099 &1.2750 &1.2823 &1.2883 \\
    IPC $\alpha_{\rm D}$     &\%                        &0.1452 &0.1453 &0.1453 &0.1430 &0.1432 &0.1428 &0.1442 \\
    $\beta_2g$               &$10^6\times$DN$^{-1}$     &2.8147 &2.8531 &1.6767 &2.8147 &2.8147 &2.8147 &2.8147 \\
    $\beta_3g^2$             &$10^{10}\times$DN$^{-2}$  &-1.0841 &-1.1119 &-0.1928 &-1.0841 &-1.0841 &-1.0841 &-1.0841 \\
    $\beta_4g^3$             &$10^{15}\times$DN$^{-3}$  &2.3025 &2.3707 &N/A &2.3025 &2.3025 &2.3025 &2.3025 \\

    $[K^{2}a + KK^I]_{0,0}$   &ppm/e     &-1.7091 &-1.7153 &-1.9086 &-1.3774 &-1.8552 &-1.7906 &-1.7044 \\
    $[K^{2}a + KK^I]_{\langle 1,0\rangle }$ &ppm/e     &0.3537 &0.3536 &0.3488 &0.3475 &0.3495 &0.3528 &0.3511 \\
    $[K^{2}a + KK^I]_{\langle 1,1\rangle }$ &ppm/e     &0.0961 &0.0960 &0.0957 &0.0980 &0.0999 &0.1010 &0.0949 \\
    $[K^{2}a + KK^I]_{\langle 2,0\rangle }$ &ppm/e     &0.0312 &0.0299 &0.0312 &0.0386 &0.0141 &0.0245 &0.0296 \\
    $[K^{2}a + KK^I]_{\langle 2,1\rangle }$ &ppm/e     &0.0090 &0.0089 &0.0090 &0.0217 &0.0064 &0.0106 &0.0097 \\
    $[K^{2}a + KK^I]_{\langle 2,2\rangle }$ &ppm/e     &0.0106 &0.0114 &0.0106 &0.0148 &0.0034 &0.0124 &0.0101 \\
    $[K^{2}a+KK^I]_{\rm H}$   &ppm/e     &0.3328 &0.3325 &0.3275 &0.3257 &0.3236 &0.3330 &0.3302 \\
    $[K^{2}a+KK^I]_{\rm V}$   &ppm/e     &0.3745 &0.3747 &0.3701 &0.3694 &0.3754 &0.3726 &0.3720 \\
    \end{tabular}
    \caption{Characterization results for SCA 20663, SCA 20828, and SCA 20829 averaged over all superpixels. Rows refer to the fiducial configuration and variants as described in section \ref{sec:characterization}. The labels lo, short, and med refer to $abcd=$ (1 3 4 6), (5 7 8 10), and (3 6 7 10) respectively. Subscripts H and V on the IPNL kernel indicate the average of the horizontal and vertical nearest neighbors to the central pixel. Subscripts in angled brackets indicate averages over nearest neighbors, averages over diagonal neighbors, etc.}
    \label{tab:char_outputs}
\end{table}

\begin{table}[]
\scriptsize  
    \centering
    \begin{tabular}{ll|ccccccc}
    \hline\hline
    Quantity     &Units &fid &128x16 &cubic &lo &short &med &$\epsilon=0.02$ \\
    \hline
    \multicolumn{9}{c}{SCA 20663} \\
    \hline
    $\sigma$(Charge, $It_{n,n+1}$)     &ke                        &0.0412 &0.0558 &0.0413 &0.0576 &0.0645 &0.0474 &0.0423 \\
    $\sigma$(Gain $g$)                 &e/DN                      &0.0469 &0.0492 &0.0474 &0.0533 &0.0547 &0.0501 &0.0465 \\
    $\sigma$(IPC $\alpha$)             &\%                        &0.0852 &0.1171 &0.0852 &0.1234 &0.1313 &0.0996 &0.0857 \\
    $\sigma$(IPC $\alpha_{\rm H}$)     &\%                        &0.0957 &0.1325 &0.0957 &0.1551 &0.1629 &0.1200 &0.0975 \\
    $\sigma$(IPC $\alpha_{\rm V}$)     &\%                        &0.1190 &0.1766 &0.1189 &0.1711 &0.1738 &0.1384 &0.1204 \\
    $\sigma$(IPC $\alpha_{\rm D}$)     &\%                        &0.0719 &0.1005 &0.0720 &0.1100 &0.1223 &0.0899 &0.0739 \\
    $\sigma(\beta_2g)$                 &$10^6\times$DN$^{-1}$     &0.3382 &0.3895 &0.1885 &0.3382 &0.3382 &0.3382 &0.3382 \\
    $\sigma(\beta_3g^2)$               &$10^{10}\times$DN$^{-2}$     &0.1647 &0.1854 &0.0398 &0.1647 &0.1647 &0.1647 &0.1647 \\
    $\sigma(\beta_4g^3)$               &$10^{15}\times$DN$^{-3}$  &0.3121 &0.3393 &N/A &0.3121 &0.3121 &0.3121 &0.3121 \\

    $\sigma([K^{2}a + KK^I]_{0,0})$   &ppm/e     &0.2937 &0.3564 &0.2852 &0.4207 &0.4924 &0.3653 &0.2794 \\
    $\sigma([K^{2}a + KK^I]_{\langle 1,0\rangle })$ &ppm/e     &0.1384 &0.1726 &0.1389 &0.2179 &0.2375 &0.1743 &0.1263 \\
    $\sigma([K^{2}a + KK^I]_{\langle 1,1\rangle })$ &ppm/e     &0.1365 &0.1692 &0.1371 &0.2012 &0.2250 &0.1699 &0.1272 \\
    $\sigma([K^{2}a + KK^I]_{\langle 2,0\rangle })$ &ppm/e     &0.1177 &0.1471 &0.1181 &0.1972 &0.2147 &0.1576 &0.1119 \\
    $\sigma([K^{2}a + KK^I]_{\langle 2,1\rangle })$ &ppm/e     &0.0999 &0.1192 &0.1002 &0.1457 &0.1599 &0.1211 &0.0910 \\
    $\sigma([K^{2}a + KK^I]_{\langle 2,2\rangle })$ &ppm/e     &0.1079 &0.1451 &0.1083 &0.1816 &0.2028 &0.1400 &0.1053 \\
    $\sigma([K^{2}a+KK^I]_{\rm H})$   &ppm/e     &0.1684 &0.2188 &0.1690 &0.2848 &0.3142 &0.2206 &0.1585 \\
    $\sigma([K^{2}a+KK^I]_{\rm V})$   &ppm/e     &0.1566 &0.2042 &0.1572 &0.2601 &0.2807 &0.1987 &0.1501 \\
    
    \hline
    \multicolumn{9}{c}{SCA 20828} \\
    \hline
    $\sigma$(Charge, $It_{n,n+1}$)     &ke                        &0.0383 &0.0828 &0.0381 &0.0594 &0.0525 &0.0426 &0.0388 \\
    $\sigma$(Gain $g$)                 &e/DN                      &0.0637 &0.0754 &0.0636 &0.0686 &0.0662 &0.0639 &0.0639 \\
    $\sigma$(IPC $\alpha$)             &\%                        &0.0846 &0.1098 &0.0845 &0.1220 &0.1149 &0.0942 &0.0858 \\
    $\sigma$(IPC $\alpha_{\rm H}$)     &\%                        &0.1046 &0.1381 &0.1046 &0.1636 &0.1528 &0.1199 &0.1057 \\
    $\sigma$(IPC $\alpha_{\rm V}$)     &\%                        &0.1317 &0.1644 &0.1316 &0.1785 &0.1763 &0.1472 &0.1345 \\
    $\sigma$(IPC $\alpha_{\rm D}$)     &\%                        &0.0637 &0.0923 &0.0637 &0.1096 &0.1099 &0.0788 &0.0669 \\
    $\sigma$($\beta_2g$)               &$10^6\times$DN$^{-1}$     &0.2383 &0.2627 &0.1432 &0.2383 &0.2383 &0.2383 &0.2383 \\
    $\sigma$($\beta_3g^2$)             &$10^{10}\times$DN$^{-2}$     &0.1343 &0.1469 &0.0313 &0.1343 &0.1343 &0.1343 &0.1343 \\
    $\sigma$($\beta_4g^3$)             &$10^{15}\times$DN$^{-3}$  &0.3030 &0.3230 &N/A &0.3030 &0.3030 &0.3030 &0.3030 \\

    $\sigma([K^{2}a + KK^I]_{0,0}$)   &ppm/e     &0.2691 &0.3256 &0.2728 &0.4999 &0.4322 &0.2966 &0.2496 \\
    $\sigma([K^{2}a + KK^I]_{\langle 1,0\rangle }$) &ppm/e     &0.1061 &0.1448 &0.1064 &0.1946 &0.2003 &0.1300 &0.1076 \\
    $\sigma([K^{2}a + KK^I]_{\langle 1,1\rangle }$) &ppm/e     &0.0975 &0.1417 &0.0979 &0.1952 &0.2079 &0.1359 &0.1008 \\
    $\sigma([K^{2}a + KK^I]_{\langle 2,0\rangle }$) &ppm/e     &0.1001 &0.1423 &0.1003 &0.1934 &0.2136 &0.1302 &0.1034 \\
    $\sigma([K^{2}a + KK^I]_{\langle 2,1\rangle }$) &ppm/e     &0.0697 &0.1030 &0.0699 &0.1359 &0.1538 &0.0969 &0.0722 \\
    $\sigma([K^{2}a + KK^I]_{\langle 2,2\rangle }$) &ppm/e     &0.0997 &0.1462 &0.0999 &0.1926 &0.2114 &0.1387 &0.1033 \\
    $\sigma([K^{2}a+KK^I]_{\rm H}$)   &ppm/e     &0.1445 &0.2053 &0.1449 &0.2624 &0.2907 &0.1915 &0.1509 \\
    $\sigma([K^{2}a+KK^I]_{\rm V}$)   &ppm/e     &0.1442 &0.1916 &0.1447 &0.2741 &0.2785 &0.1765 &0.1462 \\
    \hline
    
    \multicolumn{9}{c}{SCA 20829} \\
    \hline
    $\sigma$(Charge, $It_{n,n+1}$)     &ke                        &0.0380 &0.0489 &0.0380 &0.0549 &0.0529 &0.0422 &0.0387 \\
    $\sigma$(Gain $g$)                 &e/DN                      &0.0450 &0.0486 &0.0447 &0.0494 &0.0484 &0.0453 &0.0450 \\
    $\sigma$(IPC $\alpha$)             &\%                        &0.1181 &0.1395 &0.1181 &0.1450 &0.1435 &0.1237 &0.1176 \\
    $\sigma$(IPC $\alpha_{\rm H}$)     &\%                        &0.1256 &0.1524 &0.1255 &0.1753 &0.1694 &0.1378 &0.1258 \\
    $\sigma$(IPC $\alpha_{\rm V}$)     &\%                        &0.1632 &0.2099 &0.1631 &0.1999 &0.2036 &0.1730 &0.1634 \\
    $\sigma$(IPC $\alpha_{\rm D}$)     &\%                        &0.0645 &0.0893 &0.0644 &0.1094 &0.1031 &0.0776 &0.0673 \\
    $\sigma$($\beta_2g$)               &$10^6\times$DN$^{-1}$     &0.2905 &0.3369 &0.1517 &0.2905 &0.2905 &0.2905 &0.2905 \\
    $\sigma$($\beta_3g^2$)             &$10^{10}\times$DN$^{-2}$     &0.1697 &0.1942 &0.0363 &0.1697 &0.1697 &0.1697 &0.1697 \\
    $\sigma$($\beta_4g^3$)             &$10^{15}\times$DN$^{-3}$  &0.3777 &0.4123 &N/A &0.3777 &0.3777 &0.3777 &0.3777 \\

    $\sigma([K^{2}a + KK^I]_{0,0})$   &ppm/e     &0.1984 &0.2727 &0.1979 &0.3592 &0.3963 &0.2653 &0.2040 \\
    $\sigma([K^{2}a + KK^I]_{\langle 1,0\rangle })$ &ppm/e     &0.0930 &0.1351 &0.0934 &0.1849 &0.2067 &0.1264 &0.0964 \\
    $\sigma([K^{2}a + KK^I]_{\langle 1,1\rangle })$ &ppm/e     &0.0963 &0.1372 &0.0967 &0.1865 &0.2035 &0.1294 &0.0999 \\
    $\sigma([K^{2}a + KK^I]_{\langle 2,0\rangle })$ &ppm/e     &0.0898 &0.1345 &0.0900 &0.1782 &0.2079 &0.1316 &0.0925 \\
    $\sigma([K^{2}a + KK^I]_{\langle 2,1\rangle })$ &ppm/e     &0.0715 &0.1023 &0.0717 &0.1307 &0.1514 &0.0970 &0.0744 \\
    $\sigma([K^{2}a + KK^I]_{\langle 2,2\rangle })$ &ppm/e     &0.0966 &0.1445 &0.0968 &0.1961 &0.2125 &0.1405 &0.1008 \\
    $\sigma([K^{2}a+KK^I]_{\rm H})$   &ppm/e     &0.1328 &0.1969 &0.1334 &0.2669 &0.2951 &0.1801 &0.1366 \\
    $\sigma([K^{2}a+KK^I]_{\rm V})$   &ppm/e     &0.1307 &0.1874 &0.1313 &0.2488 &0.2873 &0.1774 &0.1362 \\
    \end{tabular}
    \caption{Standard deviations over all superpixels for the characterization results of SCA 20663, SCA 20828, and SCA 20829. Error on the mean can be obtained by dividing by 32 (or $32\sqrt{2}$, for the 128x16 configuration). See caption of table \ref{tab:char_outputs} for more details on rows and columns.}
    \label{tab:char_output_errors}
\end{table}

\begin{figure}[hbtp]
    \centering
    \includegraphics[width=6.5in]{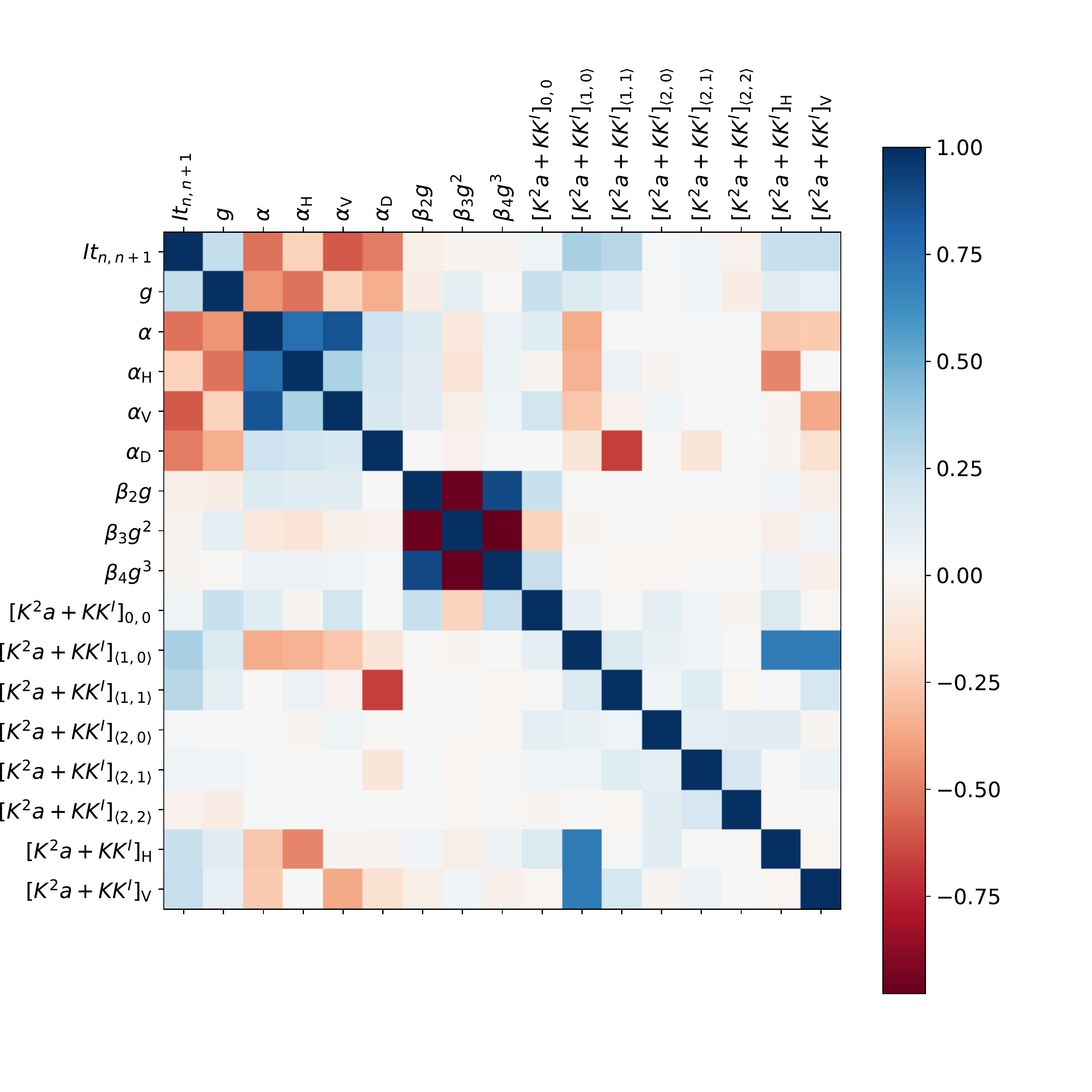}
    \caption{Correlation matrix of characterization results for SCA 20829. As explored in previous studies of IPC (e.g. \citealt{2016SPIE.9915E..2ID}), we find the IPC $\alpha$ parameters to be anticorrelated with signal level. The checkerboard pattern show by the correlations of the $\beta$ parameters arises from the non-orthogonality of the standard polynomial basis which we use to fit the CNL coefficients.}
    \label{fig:paramcorrplot}
\end{figure}

The mean values (over all superpixels) of the IPNL and CNL parameters for the three flight candidate detectors are shown in table \ref{tab:char_outputs}. Table \ref{tab:char_output_errors} shows the standard deviations of these results, and Fig. \ref{fig:paramplot} show these results in graphical form. In all three SCA cases, varying the configurations of our machinery produces little variation in the results (except, as one would expect, in the $\beta$ parameters when the degree of the CNL fit is changed), indicating the overall robustness of our method to the details of the \textsc{solid-waffle} pipeline. We note, however, that in all cases choosing timeslices clustered at the beginning of the ramp (1 3 4 6) results in a central IPNL value $[K^2a+KK^I]_{0,0}$ biased high by 0.3--0.4 ppm/e, as compared to the fiducial configuration. One would expect this effect in the presence of persistence and burn-in; however, as indicated both by Fig. \ref{fig:ph} and the absence of this effect in the (5 7 8 10) and (3 6 7 10) configurations, persistence and burn-in effects are negligible for our this measurement in our fiducial case. We also note that a smaller ($\sim 0.2$ ppm/e) difference is seen in the central IPNL value when the CNL is fit to 3rd order, indicating the significance of high orders in the CNL component of our analysis. In general, the measurement of $[K^2a+KK^I]_{0,0}$ depends on a correction for the derivative of the CNL curve, and hence is sensitive to modeling changes such as the polynomial order, or to biases from persistence or settling at the beginning of an exposure. In contrast, the ``nearest neighbor'' IPNL measurement, $[K^2a+KK^I]_{\langle 1,0\rangle}$ is much more robust: the largest difference in any of the variations from the fiducial model is 0.042, 0.013, and 0.006 ppm/e for SCAs 20663, 20828, and 20829. For SCA 20663, there is a trend that the IPNL kernel increases for time intervals centered at later times (i.e., as we go from ``lo''$\rightarrow$``fid''$\rightarrow$``med''$\rightarrow$``short''), which may suggest that the IPNL kernel in SCA 20663 is signal-dependent.

\begin{figure}
    \centering
    \includegraphics[width=6.5in]{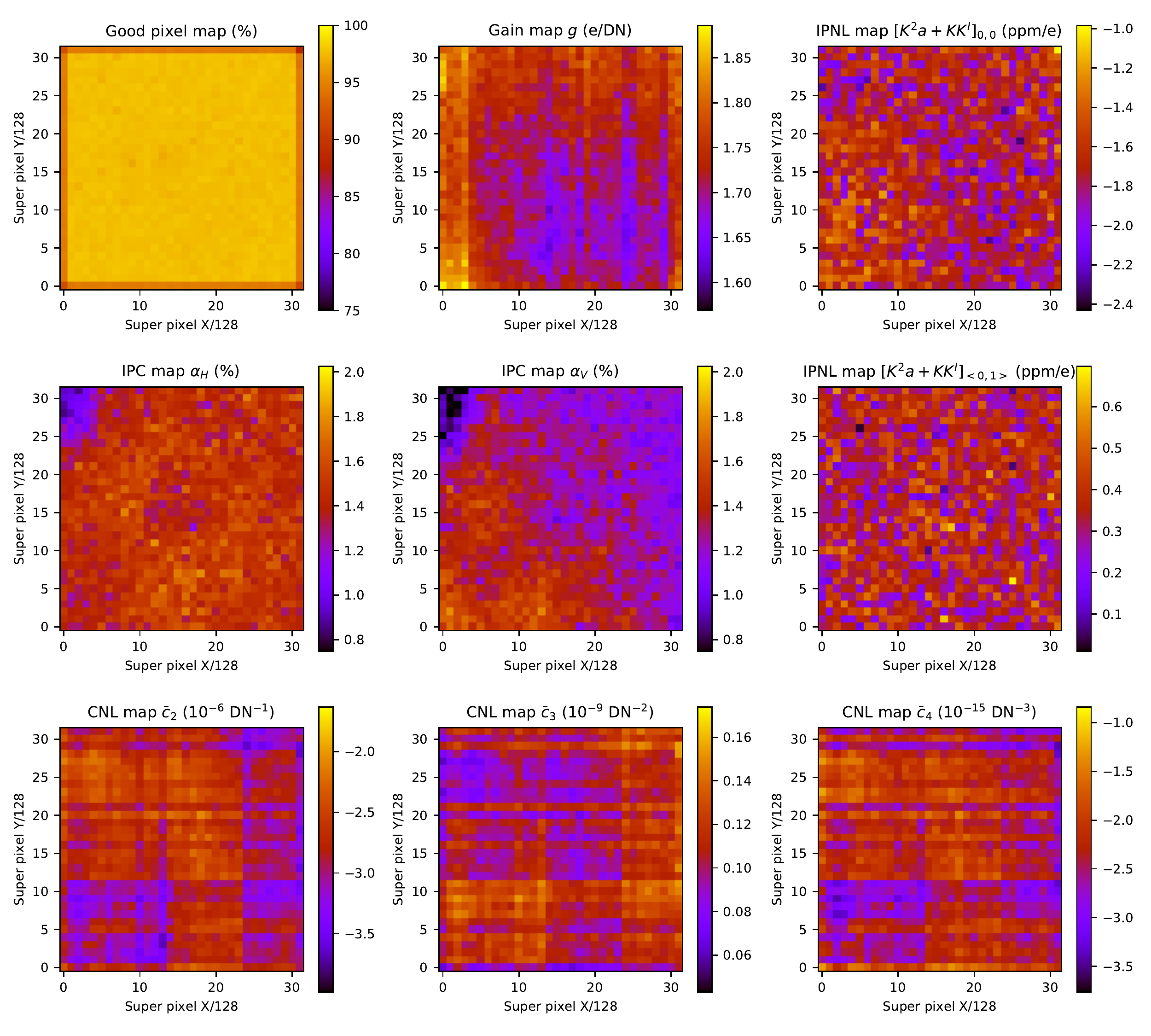}
    \caption{The spatial maps of derived quantities for SCA 20829, binned into 1024 superpixels of $128\times 128$ pixels each, for the fiducial run. Note the pronounced spatial variation of IPC in this detector. The maps of the classical non-linearity coefficients show some structure and significant anticorrelation of the polynomial components.}
    \label{fig:ninepanel}
\end{figure}

The spatial variation of quantities in the fiducial run for SCA 20829 is shown in Fig.~\ref{fig:ninepanel}.

\begin{figure}
    \centering
    \includegraphics[width=2in]{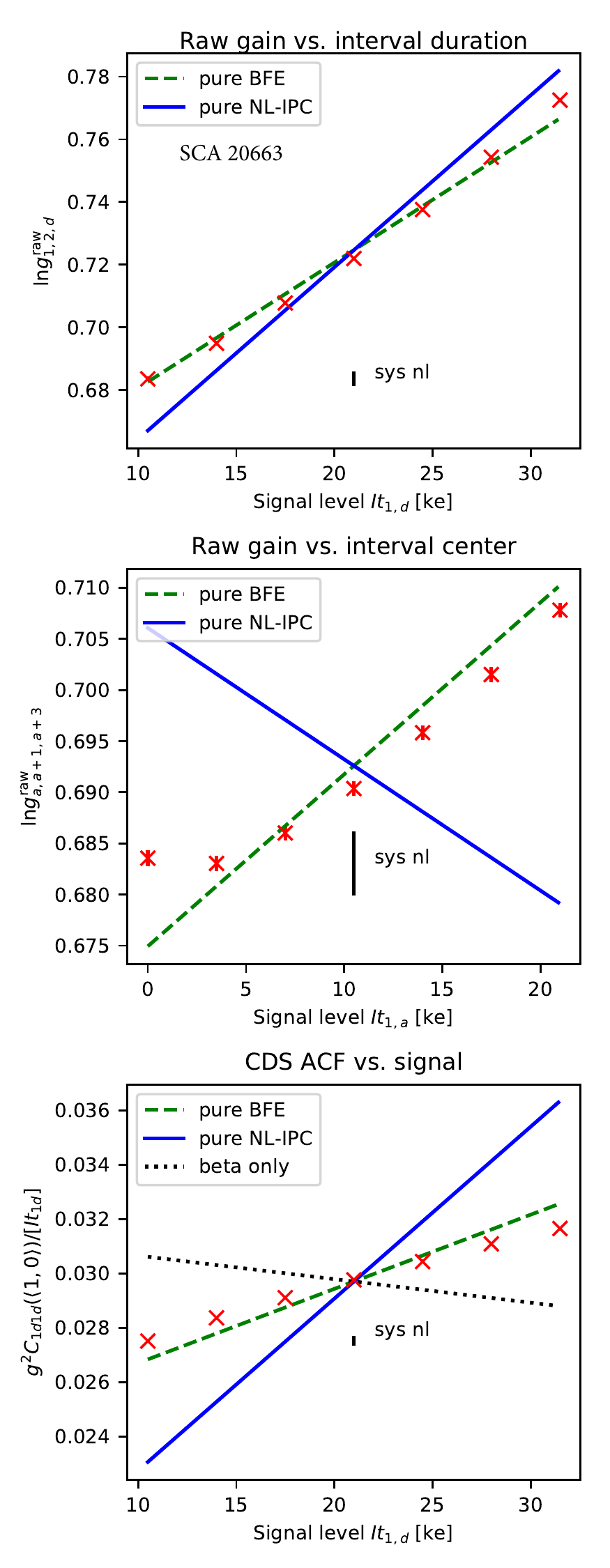}
    \includegraphics[width=2in]{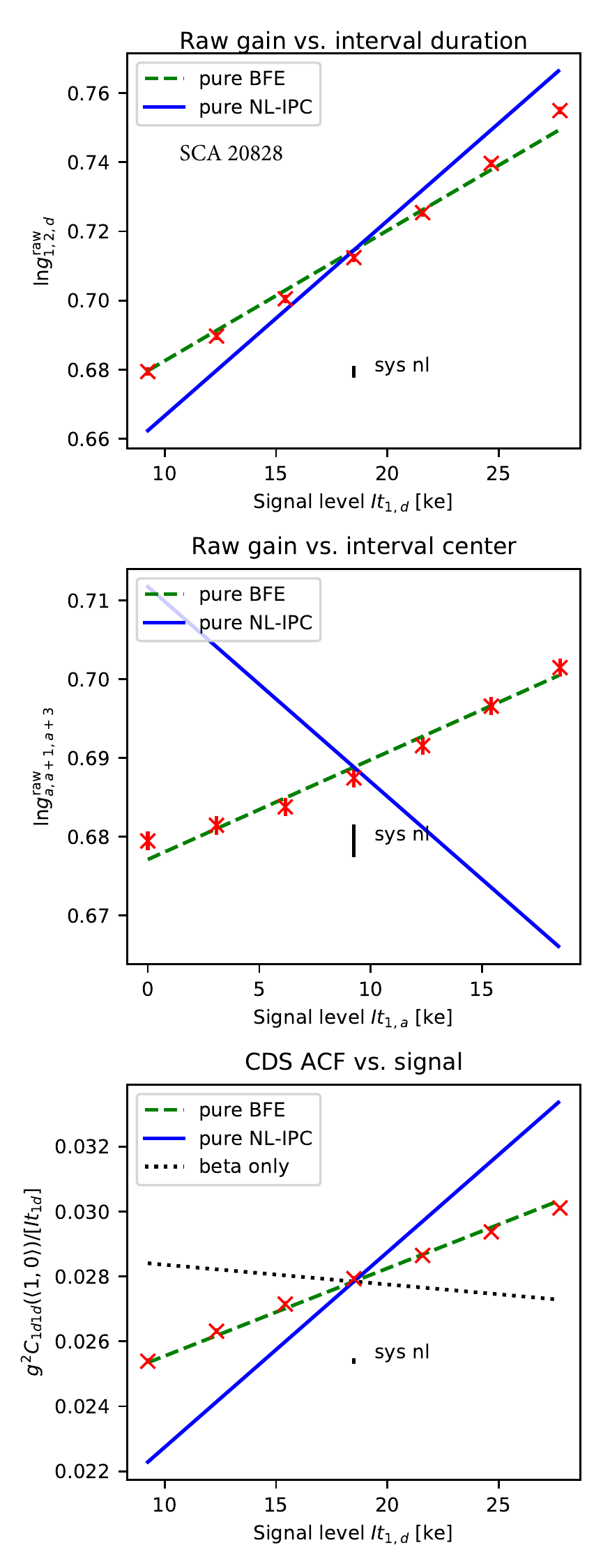}
    \includegraphics[width=2in]{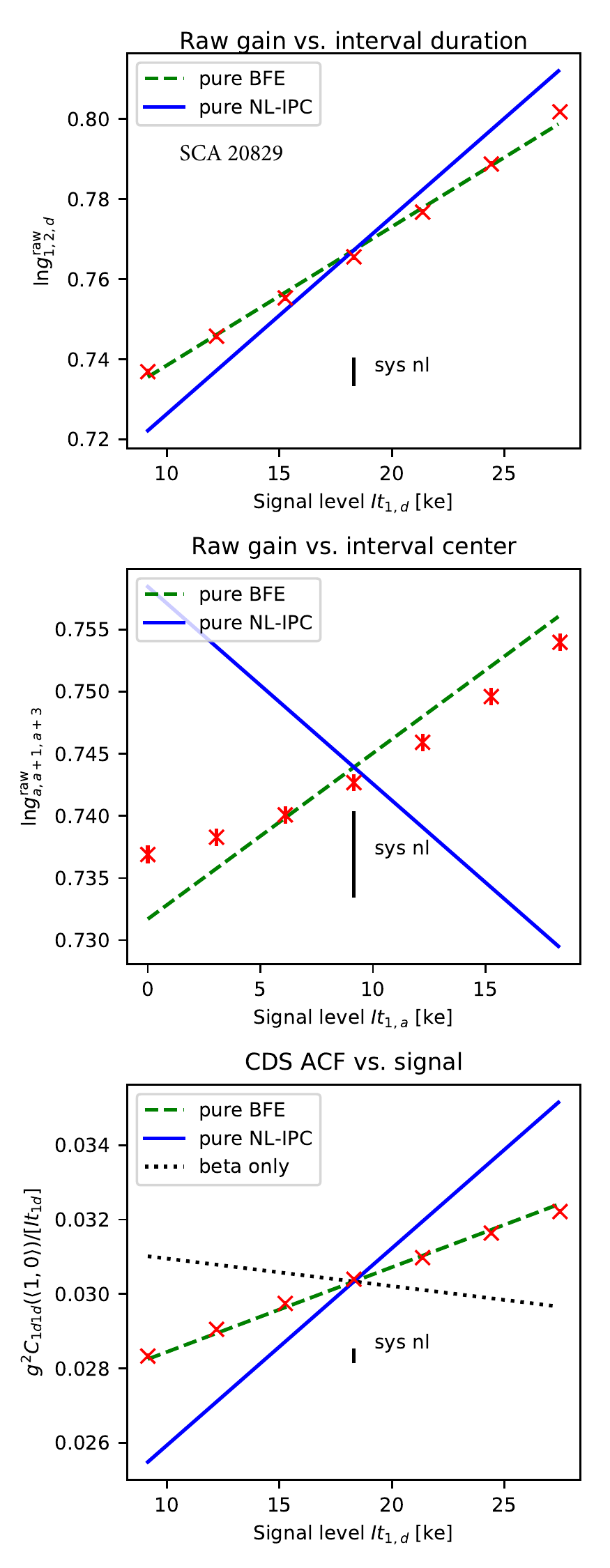}
    \caption{The dependence of raw gain and nearest-neighbor correlation functions on signal level (Methods 2 and 3 of Paper II) for the flight candidate detectors: SCA 20663 (left), SCA 20828 (middle), and SCA 20829 (right). Only the slopes, not the mean values, are meaningful. The lines show the theoretical prediction of these dependences if the IPNL is due to pure BFE (green dashed line) and pure NL-IPC (blue solid line). Just as for the development detector SCA 18237, the pure BFE provides a much better description. Note that this plot was generated with the improved model of this paper (higher-order non-linearities) turned off for consistency with Paper II.}
    \label{fig:BFEvsNLIPC}
\end{figure}

In Paper II, we used the slope of the mean-variance relation and the equal-time nearest neighbor correlation function $C_{abab}(\langle 1,0\rangle)$ to distinguish the pure BFE from the pure NL-IPC model (see Figure 6 of Paper II). We have repeated the same analysis for the flight candidate SCAs in Fig.~\ref{fig:BFEvsNLIPC}. As can be seen there, the pure BFE model is strongly preferred in all three cases for all three tests.

Finally, we compare the high-intensity flats to measurements on the low-intensity flats, also based on 1024 $128\times 128$ super-pixels. Due to the lower maximum signal level, we use only a quadratic polynomial for the classical non-linearity in the low-intensity data. We provide results here for SCA 20829 (we found similar results for the other two SCAs) using time slices 1,5,6,10 for the correlation function for IPNL determination. For SCA 20829, the mean charge accumulated was 7110 e/p in 10 frames.
For SCA 20829, the IPC measurements $\alpha_{{\rm low}\,I} - \alpha_{{\rm high}\,I}$ are $0.029\pm0.005$\% (horizontal), $0.052\pm0.005$\% (vertical), and $-0.009\pm0.003$\% (diagonal), where the error bars are ``sigma on the mean'' of the 1024 superpixels. The changes in $\alpha$ are statistically significant, although less than the difference between autocorrelation and SPR IPC measurements (see \S\ref{sss:ipc}). We find a change in gain $\ln g_{{\rm low}\,I} - \ln g_{{\rm high}\,I}$ of $0.0085\pm 0.0005$ (recall this gain is obtained from the photon transfer curve extrapolated to zero signal).

\subsubsection{Comparison to single pixel reset and hot pixel results for IPC}
\label{sss:ipc}

We have compared our autocorrelation measurements of IPC to measurements based on the single-pixel reset (SPR) method. In the SPR method \citep{2008SPIE.7021E..04S}, approximately every 8th pixel in row and in column are reset (so $\sim 1/64$ of the total pixels). The SPR data consists of 2 frames, one taken before and one taken after the reset. The data were processed in the following steps:
\begin{list}{$\bullet$}{}
\item The difference between before and after frames was taken.
\item We apply a correction for the non-linearity curve from \S\ref{ss:cnl-curve}. We use a quartic polynomial fit to the first 10 frames of flat field data from Sets 1 \& 3, or 16 flats in total. The correction is applied by numerical inversion of the polynomial.
\item For each pixel that was reset, we determine a background by selecting a $7\times 7$ enclosing region centered on that pixel, and a $3\times 3$ kernel region around that pixel. The median of the $7^2-3^2=40$ pixels in the enclosing region but not the kernel region is taken as ``background.''
\item The kernel at that pixel is estimated as $K_{\Delta i,\Delta j} = S^{\rm corr}(\Delta i,\Delta j)/\sum_{\Delta i',\Delta j'} S^{\rm corr}(\Delta i',\Delta j')$, where $S^{\rm corr}$ is the signal (following non-linearity correction and background subtraction).
\item This leads to a $3\times 3 \times 512\times 512$ array\footnote{In principle; the actual organization in Python is $9\times 512\times 512$.}, since the kernel is $3\times 3$ and is measured at every 8th pixel in the row and column directions on a $4096\times 4096$ array (recall $4096/8=512$). Some positions are missing due to reference pixels and/or readout channel edge effects; these are filled in by copying the nearest neighbor with a measurement.
\item Maps of averaged quantities such as $\alpha_{\rm H}$, $\alpha_{\rm V}$, $\alpha$, and $\alpha_{\rm D}$ can be extracted from the $3\times 3 \times 512\times 512$ master array.
\end{list}
This can be done for each SCA and for each SPR reset level (we studied 5 levels ranging from $\sim 5000$ to $\sim 17000$ DN). The lowest ($\sim 5000$ DN) level is used as default for our plots.

\begin{figure}
    \centering
    \includegraphics[width=6.5in]{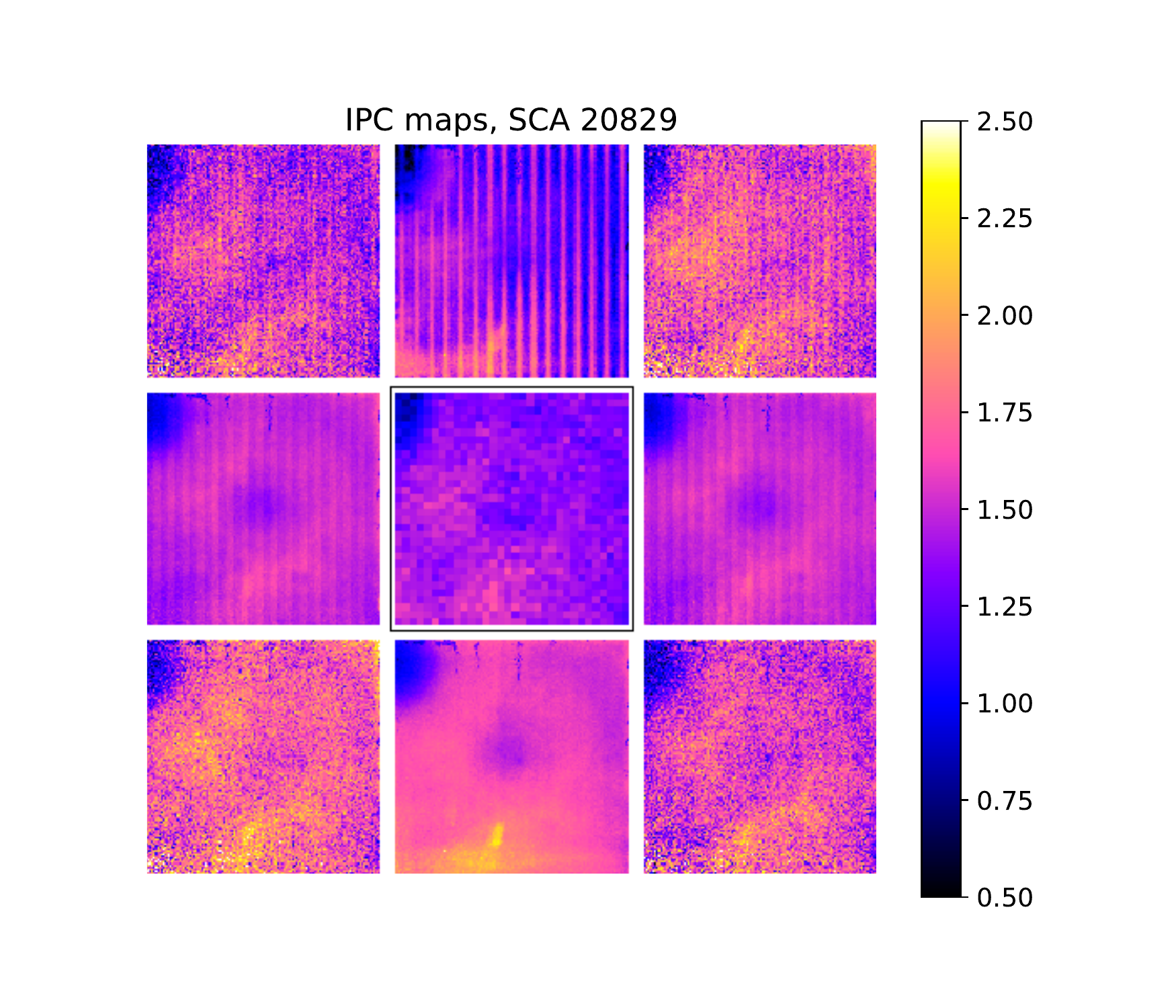}
    \caption{Maps of IPC for SCA 20829. We have shown a $3\times 3$ grid of panels. The center panel (in the black square) is $100\alpha$ as determined from the autocorrelation measurements using 16 flats. The other 8 panels show SPR measurements with $100K_{1,0}$ to the right, $100K_{0,1}$ above, $100K_{-1,0}$ to the left, $100K_{0,-1}$ below, and $1000K_{\pm1,\pm1}$ at the diagonals (note the additional factor of 10 stretch due to the diagonal IPC being lower). The autocorrelation measurements are binned into 1024 $128\times 128$ superpixels and the SPR measurements are binned into 16384 $32\times 32$ superpixels. Note that the same large-scale spatial features are present in both SPR and autocorrelation measurements. The upper panels show prominent vertical stripes due to the trailing pixel effect.}
    \label{fig:ipcmap}
\end{figure}

\begin{figure}
    \centering
    \includegraphics[width=2.1in]{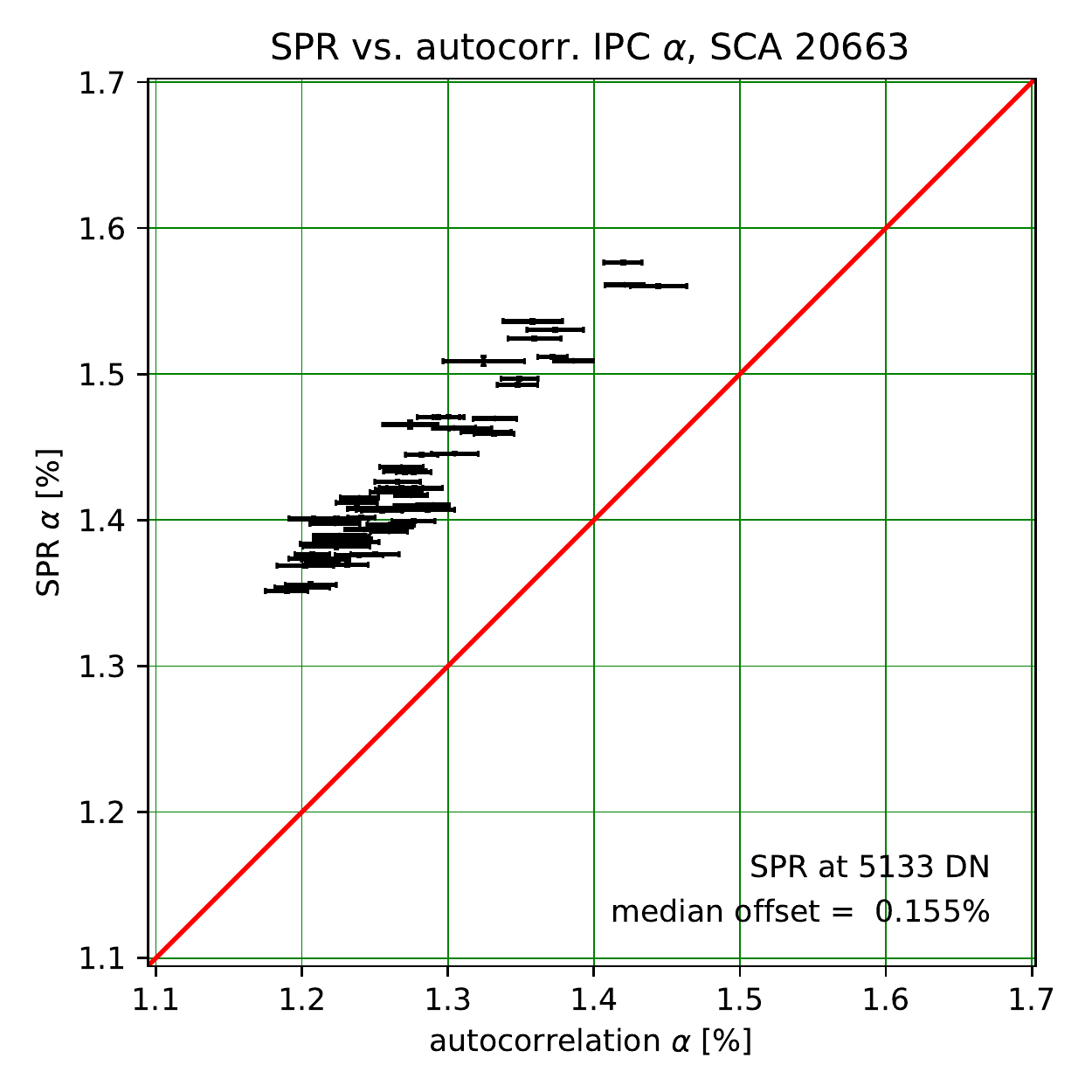}
    \includegraphics[width=2.1in]{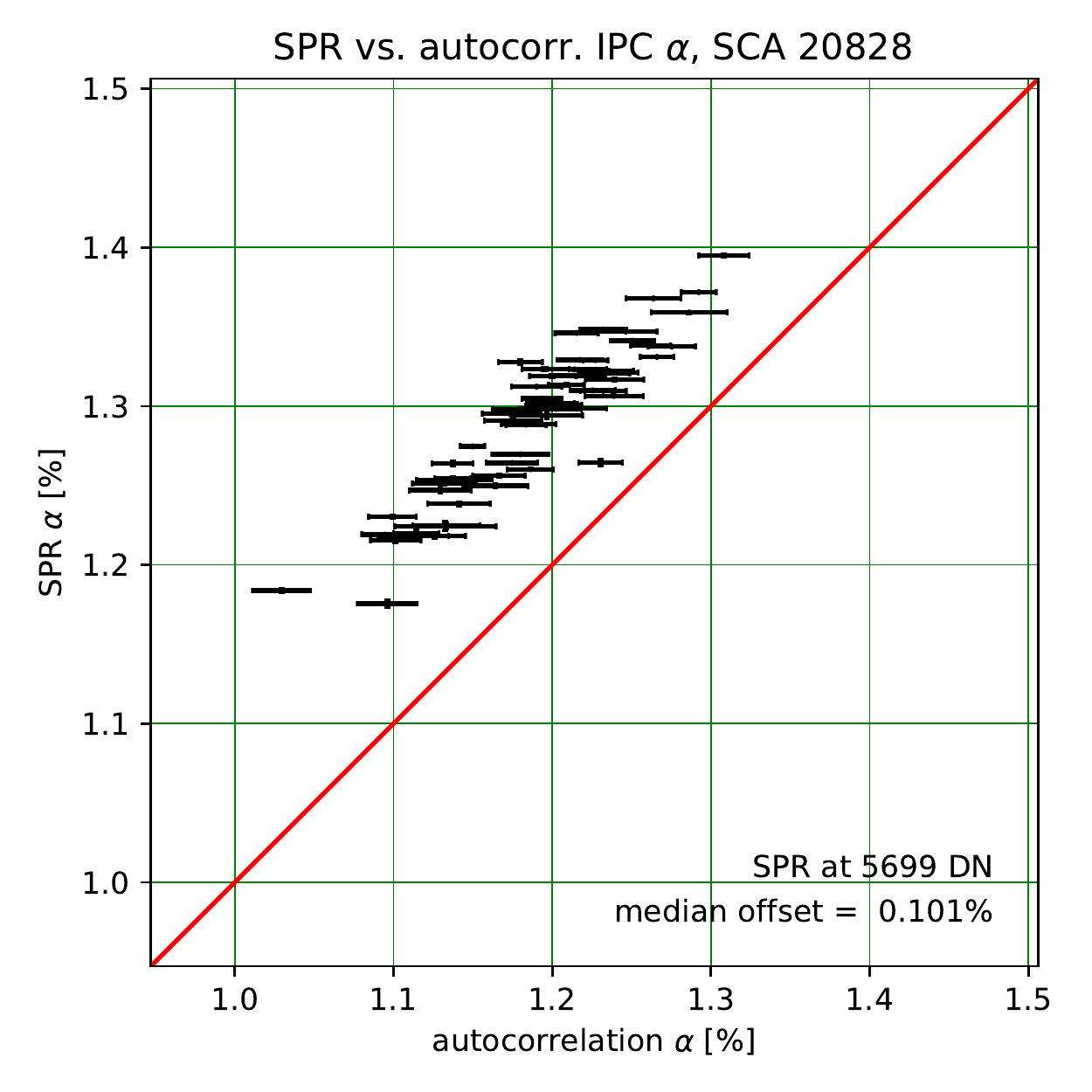}
    \includegraphics[width=2.1in]{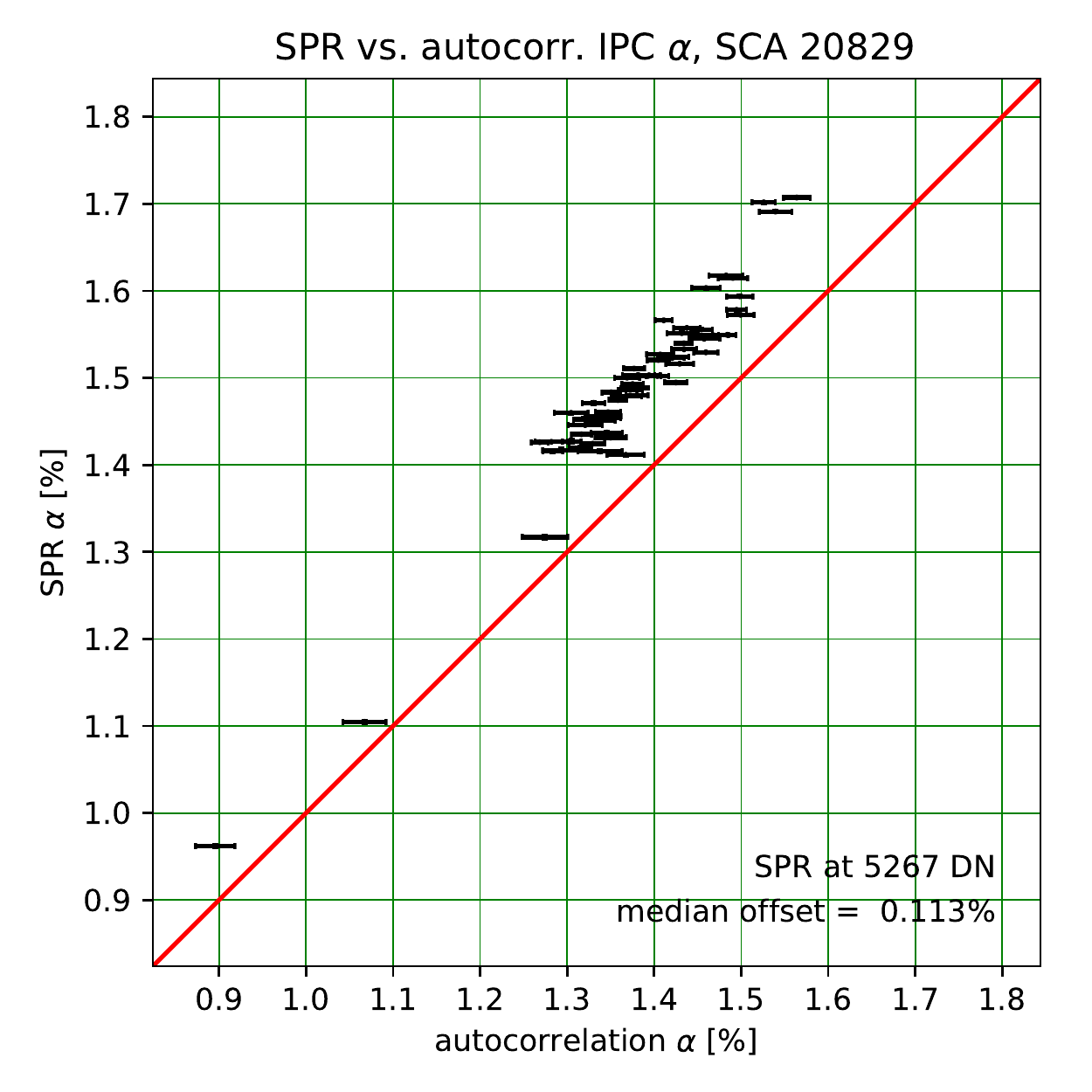}
    \caption{The comparison of $\alpha$ as determined from the single pixel reset method, versus from the autocorrelation using 16 flats (Sets 1 and 3). The data are rebinned into 64 $512\times 512$ super-pixels, each plotted as a separate point. The median central pixel level of the SPR data is indicated in DN. The median offsets ($\alpha_{\rm SPR} -\alpha_{\rm autocorr}$ is also shown; it is positive for all 3 SCAs. Note the very strong correlation between the SPR and autocorrelation results, indicating that the two methods are measuring the same spatial structure. Error bars are errors on the mean, with values outside the range 0.00--0.05 clipped.}
    \label{fig:cspr}
\end{figure}

Maps of the IPC, including comparison with the autocorrelation results from {\sc Solid-waffle}, and shown in Fig.~\ref{fig:ipcmap}. The spatial structure in the IPC is very similar in the autocorrelation measurements and the SPR measurements, although the latter have much higher S/N ratio. This can be seen quantitatively in Fig.~\ref{fig:cspr}, where we bin both results into 64 $512\times 512$ super-pixels and plot $\alpha_{\rm SPR}$ versus $\alpha_{\rm autocorr}$. There is an almost one-to-one mapping between the two, but with a systematic offset of 0.155/0.101/0.113\% for SCAs 20663/20828/20829. Such an effect was seen in the development detectors as well, e.g., in Paper I we found that for SCA 18237 there was a 0.06\% offset between the hot pixel and autocorrelation IPC measurements (again with the autocorrelation measurement being lower).

One clue to the nature of the offset can be seen in the top-center panel of Fig.~\ref{fig:ipcmap}, i.e., $K_{0,1}$, where we observe a set of 16 vertical bars across the SCA that are not present in the bottom-center panel ($K_{0,-1}$). The ``up vs.\ down'' asymmetry is not possible for DC capacitance, which is inherently symmetric whether the shapes of the conducting surfaces are symmetric or not. This ``vertical trailing pixel effect'' (VTPE) is a cross-talk effect that has been observed before in WFIRST development devices.\footnote{We thank Dave Content, Jeff Kruk, and Bernie Rauscher for presentations to the Formulation Science Working Group on this issue.} The effect traces the readout pattern (see Fig.~\ref{fig:h4rgfig}). The magnitude of the effect is thus related to the time from switching to the next row until that pixel was read. The median offsets $\alpha_{\rm SPR} - \alpha_{\rm autocorr}$ are 0.113\% (average of 4 nearest neighbors), but are 0.016\% in the horizontal direction and 0.203\% in the vertical direction, consistent with an effect primarily affecting the vertical direction.

\begin{figure}
    \centering
    \includegraphics[height=6.35in]{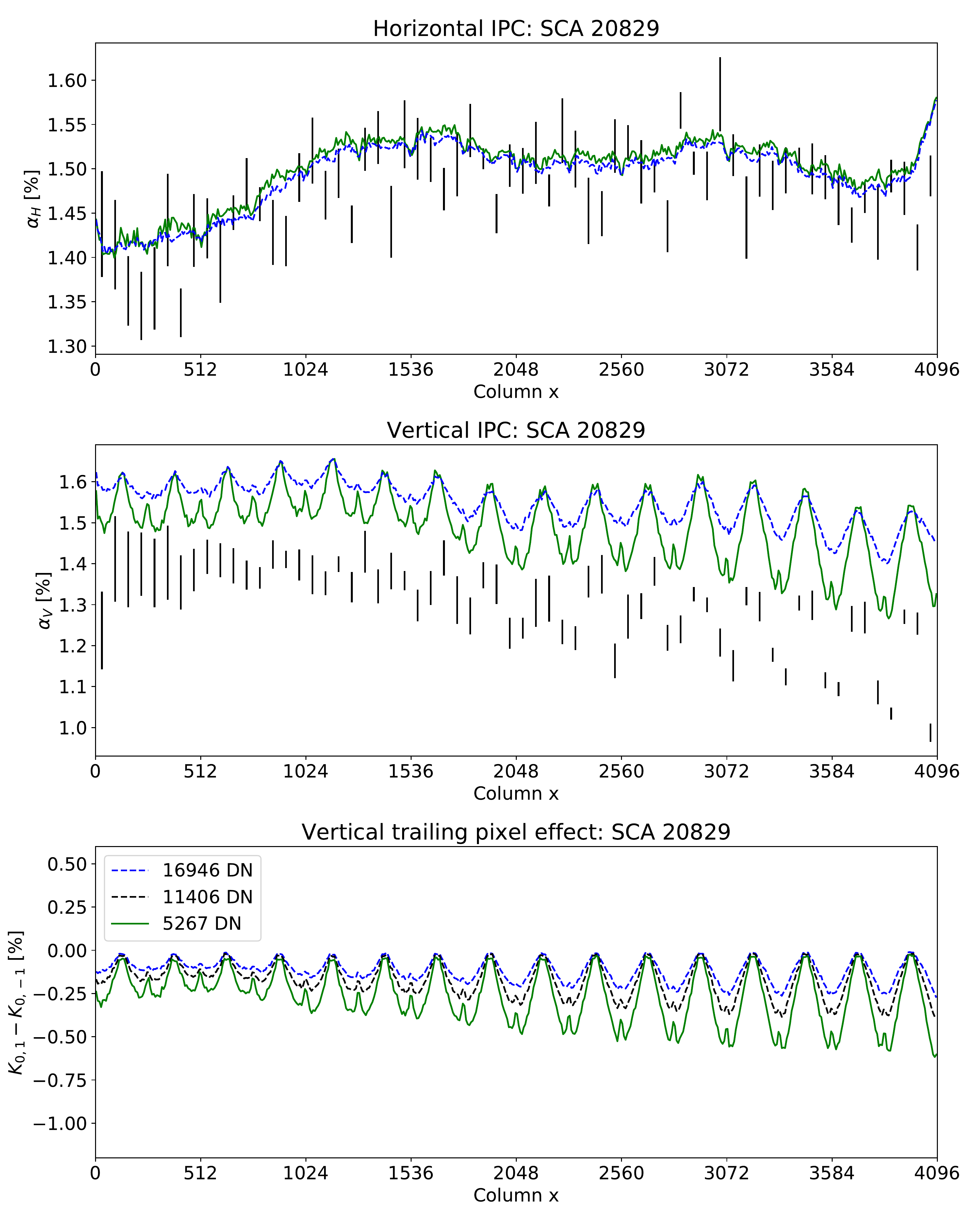}
    \caption{{\em Top panel}: The horizontal IPC $\alpha_{\rm H}$ obtained from SPR (curves: green solid at low median signal and blue dashed at high median signal), versus autocorrelations (black error bars). The data are median-filtered in each column; black points were computed in 64 column wide super-pixels. Note the generally good agreement. {\em Middle panel}: The same for vertical IPC $\alpha_{\rm V}$. The SPR measurements show the sawtooth pattern characteristic of the VTPE, and the percentage amplitude is larger at lower signal levels. The autocorrelation measurements, which are performed at very low contrast, are even lower than the green curve, and show the same sawtooth pattern. {\em Bottom panel}: The VTPE at 3 signal levels (the median contrast level between the pixel that was reset and the pixel above it is shown), as measured by $K_{0,1}-K_{0,-1}$ in the SPR data.}
    \label{fig:vtpe}
\end{figure}

This is investigated further in Fig.~\ref{fig:vtpe}, where we compare the SPR and autocorrelation measurements. In order to detect the banding pattern, here we measure autocorrelations in superpixels that are 64 columns wide and 256 rows tall (so there are two columns of superpixels per readout channel). This shows visually that the offsets between the SPR and autocorrelation measurements are mostly in $\alpha_{\rm V}$ rather than $\alpha_{\rm H}$ (top vs.\ middle panels). Moreover, the signal dependence of IPC (as measured through SPR) is much stronger in the vertical direction (top vs.\ middle panel), and the VTPE is stronger (i.e., more negative) as a percentage at lower signal levels (bottom panel). We suspect that the autocorrelation result -- which is obtained at very low contrast -- is due to the VTPE becoming even larger (in a percentage sense) at these low contrasts.

\begin{figure}
    \centering
    \includegraphics[width=4.5in]{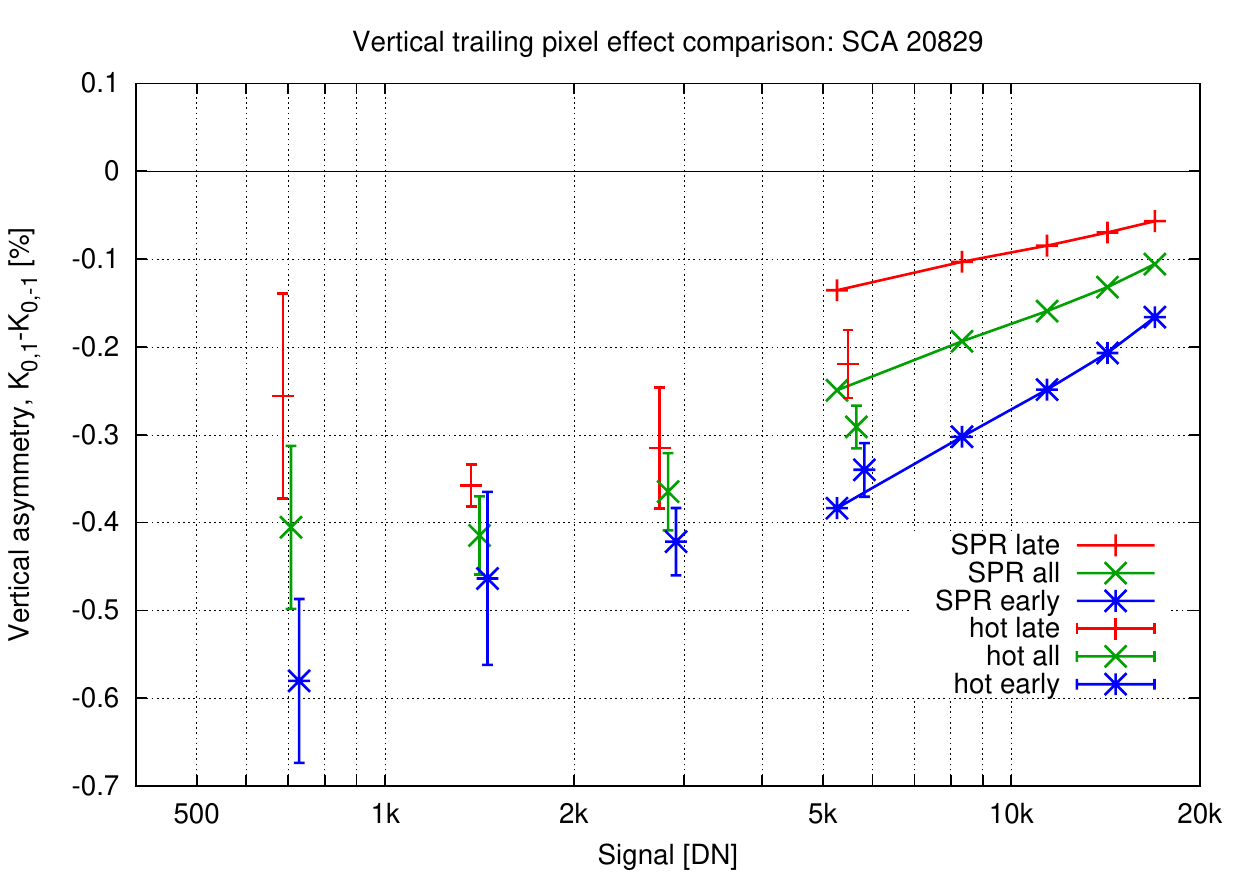}
    \caption{The vertical asymmetry measured by single pixel resets (points with lines connecting them) and hot pixels (points with error bars). We show both results for all pixels (green), and split into ``early'' pixels (blue) and ``late'' pixels (red), depending on whether the pixel is in the first 64 or last 64 of the 128 pixels in that row read in that readout channel. Note that (i) the magnitude of the effect is larger at lower signal levels, and this trend continues at the lower signal levels explored by the hot pixel data; and (ii) the ``early'' pixels show a systematically larger (in magnitude) effect than the ``late'' pixels in both measurements.}
    \label{fig:vtpe_hot}
\end{figure}

One more test of the signal dependence of VTPE is shown in Fig.~\ref{fig:vtpe_hot}, where we include both the single pixel reset data and measurements from hot pixels (see Paper II, \S5.4), which continue down to lower signal levels. The hot pixels were obtained from the dark exposures: there are 54 such exposures in our sequence (excluding first darks following a flat, which are affected by persistence). The pixels were grouped into signal levels with a width of a factor of 2 in signal (500--1k DN, 1k--2k DN, 2k--4k DN, and 4k--8k DN); we show the mean and standard deviation of the asymmetry measured across 6 runs of {\sc Solid-waffle} each with $54/6=9$ darks. We see that the tendency for the asymmetry to get stronger continues toward lower signal levels. Moreover, we have split the pixels into ``early'' pixels (the first 64 read in their row in their readout channel; these correspond to the valleys in Fig.~\ref{fig:vtpe}) and ``late'' pixels (the last 64 read in their row in their readout channel; these correspond to the peaks in Fig.~\ref{fig:vtpe}). While the S/N ratio of the hot pixel measurement is low, the peak-valley pattern is also present.

\section{Discussion}
\label{sec:discussion}

This paper has presented a treatment of correlations in infrared detector array flat fields in Fourier space, analogous to the treatment of CCDs by \citet{2019A&A...629A..36A}, that enables us to work to all orders in the BFE. We have updated the {\sc Solid-waffle} analysis framework to make use of this new model for the correlation function. We have tested our implementation of the Fourier space formalism on simulations, finding that the output BFE central and averaged nearest neighbor coefficient values match the inputs to within 1\% when also including classical non-linearity polynomial fits up to the quartic coefficient.  This demonstrates a significant performance improvement over the previous Paper I biases of 12.1\% and 2.7\% for the recovered central and nearest neighbor BFE coefficients and supports the hypothesis that the biases were caused by lack of higher order terms in the configuration space analysis.

We also ran our updated analysis on the SCA 18237 data presented in Paper II.  As the main focus of this paper is on the three flight candidate detectors, we omit lengthy discussion of SCA 18237 but do note one new observation pertaining to the behavior of the nearest neighbor IPNL.  In Paper II, we found the nearest neighbor IPNL coefficient appeared to significantly decrease as the choice of time baseline increased.  Here, we find the difference between the longer and shorter baseline results to be smaller (0.0169 vs.\ 0.0274 ppm/e), less significant, and go in the opposite sense where the coefficient now increases with longer time baselines.

We have also run our analyses on the first three flight candidate detector arrays (SCAs 20663, 20828, and 20829), and presented most of our analyses for SCA 20829. We observe persistence and burn-in (excess signal in a 2nd flat exposure relative to the 1st). These effects are spatially correlated and generally similar to other examples we have seen during the {\slshape WFIRST} development program. As with development detector SCA 18237, the BFE dominates over NL-IPC in all three flight candidates. The fiducial measurements of the IPNL central pixel for each candidate are larger than that of SCA 18237, which also has a lower value of $\alpha$. This indicates that the BFE is stronger in the flight candidates than in SCA 18237. Paper II calculates that for an idealized PSF measurement, the BFE in the development detector induces an effective central pixel area decrease of $\sim 2\%$, depending on band. Given the comparative IPNL and IPC values, we expect the flight candidates to exhibit a slightly large effect of the BFE on PSF, although of a similar order of magnitude. See the discussion of Paper II for more details on the PSF model and statistical requirements.

Across all configuration settings, SCA 20828 has the lowest (or is tied for having the lowest) values of inter-pixel capacitance $\alpha_{H}$ and $\alpha_{D}$, though there is spatial variation for all SCAs and the ranges of $\alpha$ have some region of overlap. When we performed the classical non-linearity fits with a cubic instead of a quartic polynomial, the coefficients $\beta_{2}g$ and $\beta_{3}g^{2}$ changed significantly, which is expected given that the standard polynomial basis is not orthogonal (in future work, we may follow \citealt{2019JATIS...5b8001R} and use the Legendre polynomial basis for this reason). This disadvantage of the standard basis is further demonstrated in Fig. \ref{fig:paramcorrplot}, which shows strong anti-correlation among odd and even $\beta$ coefficients.

The measurement of the central pixel IPNL kernel from the flat fields in the flat/dark sequence data -- i.e., $[K^2a+KK^I]_{0,0}$, which describes how the charge level in a pixel affects its response -- is biased if only the first few frames of data are used. As discussed in section \ref{ss:application}, this particular measurement is sensitive to the slope of the non-linearity curve, and therefore is more sensitive to systematic errors in the non-linearity curve caused by the model (polynomial order), and by persistence and burn-in that make the non-linearity curve for each exposure slightly different. We believe the latter was exacerbated by allowing the detector to sit in the saturated state for $\sim 90$ s repeatedly during the test, and the lamp intensity for flat/dark testing for future SCAs has been reduced to mitigate this effect. The ``nearest neighbor'' measurement $[K^2a+KK^I]_{\langle 1,0\rangle}$, where we have placed a calibration requirement, is much less affected; maximum variations from the fiducial choice of time intervals are 0.042, 0.013, and 0.006 ppm/e for SCAs 20663, 20828, and 20829. The variation in SCA 20663 shows a systematic trend with signal level that may hint at signal-dependent BFE.

The inter-pixel nonlinearity is also observed to be nearly but not exactly symmetric between rows and columns: the array-averaged horizontal minus vertical asymmetry for the fiducial measurement is $([K^2a]_{\rm H}-[K^2a]_{\rm V})/([K^2a]_{\rm H}+[K^2a]_{\rm V}) = -0.015\pm 0.008$ (SCA 20663), $-0.046\pm 0.008$ (SCA 20828), and $-0.059\pm 0.008$ (SCA 20829) based on Table~\ref{tab:char_outputs}. This is a similar order of magnitude to what we observed in Paper II for SCA 18237, but of the opposite sign; it is also clear there is some variation from one detector array to another. We plan to investigate this asymmetry further after developing a better model for the vertical trailing pixel effect. Finally, there is a rapid fall-off of the IPNL kernel with distance in the flight candidate SCAs (though not quite as fast as for SCA 18237): the effect on the diagonal neighbors is less than the nearest neighbors by a factor of $[K^2a]_{\langle 1,1\rangle}/[K^2a]_{\langle 1,0\rangle}= $0.27--0.35, and for the second-nearest neighbors this factor is $[K^2a]_{\langle 2,0\rangle}/[K^2a]_{\langle 1,0\rangle}= $0.07--0.17.

We have compared the IPC maps from flat field autocorrelation measurements to those from the single pixel reset method. In general the agreement is good: the same spatial structures are seen in both methods, and the median difference is 0.113\% (in SCA 20829), with the SPR measurement of $\alpha$ being systematically larger in all of the SCAs. The discrepancy is larger in the vertical direction, which we have attributed to the vertical trailing pixel effect (VTPE) -- a non-linear effect that results in a lower signal in the pixel immediately above a pixel containing more electrons. The VTPE has a spatial structure that traces the readout pattern.
The VTPE will have to be calibrated for weak lensing applications because (i) it is asymmetric in vertical vs.\ horizontal directions, thus impacting ellipticity measurements; and (ii) it is non-linear: it is a larger effect (in magnitude) for the faint galaxies used for science measurements than for bright stars used to determine the PSF. For {\slshape WFIRST}, the requirement on PSF ellipticity knowledge is $5.7\times 10^{-4}$ RMS per component ($e_1$ or $e_2$, in the convention of \citealt{2002AJ....123..583B}). Image simulations have shown that in $J$-band (which is the most affected shape measurement band on WFIRST), $\partial e_1/\partial \alpha_{\rm V} = -2.01$ and $\partial e_2/\partial \alpha_{\rm V} = -0.05$.\footnote{See Table 4 of \citet{2016PASP..128i5001K}; note that $\partial e_i/\partial \alpha_{\rm V}$ is given by $(S_{i,\alpha} - S_{i,+})/2$ in their notation.} This means that the entire $4.7\times 10^{-4}$ RMS per component budget is saturated if we have a remaining difference after calibration in $\alpha_{\rm V}$ between the PSF stars and the faint galaxies used for shape measurement of 0.040\% (i.e., $5.7\times 10^{-4}/\sqrt{(2.01^2+0.05^2)/2}$). In practice the VTPE will have to be a sub-allocation of this budget. Since we found a difference of 0.2\% in $\alpha_{\rm V}$ between the high contrast single pixel reset measurement and the low contrast autocorrelation measurement of $\alpha_{\rm V}$, a correction for VTPE will be required for {\slshape WFIRST}. The VTPE may also have to be corrected for precision astrometry with {\slshape WFIRST} \citep[e.g.][]{2015JKAS...48...93G, 2017arXiv171205420S, 2018AJ....155..102M, 2019BAAS...51c.211G} -- objects that appear in columns near the ``valleys'' in the lower panel of Fig.~\ref{fig:vtpe} will appear displaced downward by a few thousandths of a pixel (i.e., a few hundred $\mu$as), and this effect will be larger for fainter objects.

Fortunately, there are a number of mitigations available for VTPE. We are planning further laboratory tests to characterize cross-talk non-linearity in the H4RG-10 detectors. A test using a projected array of spots on an H4RG-10 is planned to be carried out at the Caltech/JPL lab (in a modification of the setup used in \citealt{2018PASP..130f5004P}), which will constrain how signal-dependent VTPE manifests itself in spot images at comparable intensities and undersampling factors to PSF stars. Since the VTPE has a specific spatial pattern and its effect is primarily in $e_1$ rather than $e_2$ in SCA-fixed coordinates, the cross-linked observing strategy of {\slshape WFIRST} with observations at multiple roll angles \citep{2015arXiv150303757S, 2019arXiv191209481T} will enable us to distinguish VTPE from a sky-fixed astrophysical signal. Furthermore, only the large scale features in VTPE, and not the ``sawtooth'' pattern of Fig.~\ref{fig:vtpe}, falls under the requirement: the sawtooth itself has a wavelength of $\Delta\theta = 256\,$pix$\,=1.37\times 10^{-4}$ radians, and hence corresponds to an angular scale $\ell = 2\pi/\Delta\theta = 46000$, well outside the range of Fourier modes used for the weak lensing cosmology program. We are in the process of reviewing our calibration procedures to determine whether additional tests are needed to characterize the VTPE and build a model to include in image simulation tools such as {\sc GalSim} \citep{2015A&C....10..121R, 2019arXiv191209481T}.

This analysis is one step toward calibrating the many non-linearity and cross-talk effects that occur in {\slshape WFIRST} detectors. We have addressed the biggest limitation of the formalism of Papers I and II by computing the correlation function $C_{abcd}(\Delta x_1,\Delta x_2)$ to all orders in the BFE, including its interaction with IPC and classical non-linearity, and allowing for non-linearity polynomials of arbitrary order. We have also applied the correlation function formalism to a larger sample of SCAs (and not just the single development detector used in Paper II), including the first three {\slshape WFIRST} flight candidates. This illustrates the power of acceptance test data as a proving ground for calibration techniques, although we again caution that the data here were taken under laboratory conditions with a laboratory controller and some properties may be different in flight. Our plans for future work now include more investigation of the vertical trailing pixel effect, as well as comparison to focused spot and speckle fringe illumination data to explore how the BFE operates at higher contrast, as will occur when we observe stars (whether for weak lensing PSF determination, flux calibration for supernovae, or microlensing sources). We will also continue to analyze the test data for additional flight candidate detectors, since at the level of precision required for {\slshape WFIRST} each detector is unique.

\section*{Acknowledgements}

We thank Eric Huff, Andr\'es Plazas, Bernard Rauscher, and Chaz Shapiro for helpful discussions and feedback. Computations for this paper were carried out at the \citet{OhioSupercomputerCenter1987}. This paper is based on data acquired at the Detector Characterization Laboratory at NASA Goddard Space Flight Center. We thank Roger Foltz, Chris Merchant, Augustyn Waczynski, and Yiting Wen for their contributions to this data set.

AC, JF, JG, and CMH acknowledge support from NASA award 15-WFIRST15-0008; Simons Foundation award 60052667; and US Department of Energy award DE-SC0019083.

{\em Software:} Astropy \citep{2013A&A...558A..33A,2018AJ....156..123A}, fitsio \citep{fitsio}, Matplotlib \citep{Hunter:2007}, NumPy \citep{numpy}, SciPy \citep{scipy}

\bibliographystyle{aasjournal}
\bibliography{main}

\end{document}